\def\spose#1{\hbox to 0pt{#1\hss}}
\def\approxlt{\mathrel{\spose{\lower 3pt\hbox{$\sim$}}
        \raise 2.0pt\hbox{$<$}}}
\def\approxgt{\mathrel{\spose{\lower 3pt\hbox{$\sim$}}
        \raise 2.0pt\hbox{$>$}}}

\def\multleft#1{\hbox to size{\vbox {\halign {\lft{##}\cr #1}}\hfill}\par}
\def\multright#1{\hbox to size{\vbox {\halign {\rt{##}\cr #1}}\hfill}\par}

\def\degmark{^\circ}
\def\boxit#1{\vbox{\hrule\hbox{\vrule\kern3pt\vbox{\kern3pt
          #1 \kern3pt}\kern3pt\vrule}\hrule}}

\def\cm{{\rm\thinspace cm}}

\def\erg{{\rm\thinspace erg}}

\def\K{{\rm\thinspace K}}
\def\keV{{\rm\thinspace keV}}
\def\km{{\rm\thinspace km}}
\def\kpc{{\rm\thinspace kpc}}

\def\pc{{\rm\thinspace pc}}
\def\ph{{\rm\thinspace ph}}
\def\s{{\rm\thinspace s}}

\def\chisq{\hbox{$\chi^2$}}

\def\cmsq{\hbox{$\cm^2\,$}}

\def\pcmcu{\hbox{$\cm^{-3}\,$}}

\def\ergpcmsqps{\hbox{$\erg\cm^{-2}\s^{-1}\,$}}
\def\ergpcmsqpspA{\hbox{$\erg\cm^{-2}\s^{-1}$\AA$^{-1}\,$}}
\def\ergps{\hbox{$\erg\s^{-1}\,$}}

\def\kmps{\hbox{$\km\s^{-1}\,$}}

\def\pcmsq{\hbox{$\cm^{-2}\,$}}

\def\phpcmsqps{\hbox{$\ph\cm^{-2}\s^{-1}\,$}}

\def\ps{\hbox{$\s^{-1}\,$}}

\documentclass[preprint]{aastex}

\received{}
\accepted{}

\slugcomment{To appear in The Astrophysical Journal v564 n1 January 1, 2002}

\shorttitle{A \emph{Chandra} X-ray Study of Cygnus A --- II. The Nucleus}
\shortauthors{Young, Wilson, Terashima, Arnaud \& Smith}

\begin{document}

\title{A \emph{Chandra} X-ray Study of Cygnus A --- II. The Nucleus}

\author{Andrew J. Young, Andrew S. Wilson\altaffilmark{1}, Yuichi
Terashima\altaffilmark{2}}

\affil{Astronomy Department, University of Maryland, College Park, MD
	20742}

\author{Keith A. Arnaud\altaffilmark{3}}

\affil{Laboratory for High Energy Astrophysics, NASA Goddard Space
	Flight Center, Greenbelt, MD 20771}

\and

\author{David  A. Smith}

\affil{Astronomy Department, University of Maryland, College Park, MD 20742}


\altaffiltext{1}{Adjunct Astronomer, Space Telescope Science
	Institute, 3700 San Martin Drive, Baltimore, MD 21218}

\altaffiltext{2}{Institute of Space and Astronautical Science, 3-1-1
	Yoshinodai, Sagamihara, Kanagawa 229-8510, Japan}

\altaffiltext{3}{also Astronomy Department, University of Maryland,
	College Park, MD 20742}


\begin{abstract}

  We report \emph{Chandra} ACIS and quasi-simultaneous \emph{RXTE} observations
  of the nearby, powerful radio galaxy Cygnus A, with the present paper
  focusing on the properties of the active nucleus. In the \emph{Chandra}
  observation, the hard ($>$ a few keV) X-ray emission is spatially unresolved
  with a size $\approxlt 1\arcsec$ (1.5 kpc, $H_0 = 50$ km s$^{-1}$ Mpc$^{-1}$)
  and coincides with the radio and near infrared nuclei. In contrast, the soft
  ($< 2 \keV$) emission exhibits a bi-polar nebulosity that aligns with the
  optical bi-polar continuum and emission-line structures and approximately
  with the radio jet. In particular, the soft X-ray emission corresponds very
  well with the [\ion{O}{3}] $\lambda 5007$ and H$\alpha$ + [\ion{N}{2}]
  $\lambda\lambda 6548$, 6583 nebulosity imaged with HST. At the location of
  the nucleus there is only weak soft X-ray emission, an effect that may be
  intrinsic or result from a dust lane that crosses the nucleus perpendicular
  to the source axis. The spectra of the various X-ray components have been
  obtained by simultaneous fits to the 6 detectors. The compact nucleus is
  detected to 100 keV and is well described by a heavily absorbed power law
  spectrum with $\Gamma_h = 1.52^{+0.12}_{-0.12}$ (similar to other narrow line
  radio galaxies) and equivalent hydrogen column $N_H ({\rm nuc}) =
  2.0^{+0.1}_{-0.2} \times 10^{23} \pcmsq$. This column is compatible with the
  dust obscuration to the near infrared source for a normal gas to dust ratio.
  The soft ($<2 \keV$) emission from the nucleus may be described by a power
  law spectrum with the same index (i.e. $\Gamma_l = \Gamma_h$), though direct
  fits suggests a slightly larger value for $\Gamma_l$. Narrow emission lines
  from highly ionized neon and silicon, as well as a ``neutral'' Fe K$\alpha$
  line, are detected in the nucleus and its vicinity ($r \approxlt 2 \kpc$).
  The equivalent width (EW) of the Fe K$\alpha$ line ($182^{+40}_{-54}$ eV) is
  in good agreement with theoretical predictions for the EW versus $N_H ({\rm
  nuc})$ relationship in various geometries. An Fe K edge is also seen. The
  \emph{RXTE} observations indicate a temperature of $kT = 6.9^{+0.3}_{-1.0}$
  keV for the cluster gas (discussed in Paper III of this series) and cluster
  emission lines of Fe K$\alpha$ and Fe K$\beta$ and / or Ni K$\alpha$.

  We consider the possibility that the extended soft X-ray emission is
  electron-scattered nuclear radiation. Given that 1\% of the unabsorbed 2 --
  10 keV nuclear radiation would have to be scattered, the necessary gas column
  ($N_H ({\rm scattering}) \simeq 3.5 \times 10^{22} \pcmsq$) would absorb the
  X-rays rather than scatter them if the gas is cold. Thus the scattering
  plasma must be highly ionized. If this ionization is achieved through
  photoionization by the nucleus, the ionization parameter $\xi > 1$ erg cm
  s$^{-1}$ and the electron density $n_e \simeq 6 \pcmcu$ given the observed
  distance of the soft X-ray emission from the nucleus. The electron column
  density inferred from the X-ray observations is much too low to account for
  the extended optical scattered light, strongly suggesting that the polarized
  optical light is scattered by dust. The presence of highly ionized Ne lines
  in the soft X-ray spectrum requires $20 \approxlt \xi \approxlt 300$ erg cm
  s$^{-1}$; these lines may originate closer to the nucleus than the extended
  soft continuum or in a lower density gas. A collisionally-ionized thermal
  model of the extended soft X-rays cannot be ruled out, but is unattractive in
  view of the low metal abundance required ($Z = 0.03 Z_\odot$).

  The hard X-ray to far-infrared ratio for the nucleus of Cygnus A is similar
  to that seen in Seyfert 1 and unobscured radio galaxies. By means of the
  correlation between hard X-ray luminosity and nuclear optical absolute
  magnitude for these classes of object, we estimate $M_B = -22.4$ for Cygnus
  A, near the borderline between Seyferts and QSOs.

\end{abstract}




%

\section{Introduction} \label{sec:intro}

Cygnus A (3C 405) is the closest and best studied of the powerful ``classical
double'' (FR II) radio galaxies. At a heliocentric redshift of $z = 0.0562$
(Stockton, Ridgway \& Lilly 1994) it is the most powerful radio source in the
3C catalog out to $z \simeq 1$, and about 1.5 orders of magnitude more luminous
than any other 3C source within $z \le 0.1$. It is interesting for a number of
reasons, including: i) radio ``hot spots'', the shocked jet material at the
termini of the jets, were first observed in Cygnus A (Hargrave \& Ryle 1974);
ii) the galaxy may be at the center of a cooling flow in a cluster of galaxies
(Arnaud et al. 1984; Reynolds \& Fabian 1996), and iii) there is evidence for a
``buried quasar'' in the nucleus (Djorgovski et al. 1991; Ward et al. 1991;
Antonucci, Hurt \& Kinney 1994; Ueno et al. 1994; Ogle et al. 1997). In this
paper we present new \emph{Chandra} X-ray observations of the nucleus and
circumnuclear environment of Cygnus A.

In the past there has been debate concerning the nature of the nucleus and
circumnuclear regions (approximately the central $4\arcsec$ [6 kpc]) of Cygnus
A, in particular the central bi-polar optical morphology. Suggested
explanations have included i) the merger of two galaxies (Baade \& Minkowski
1954), ii) the intersection of a single galaxy by a dust lane (Osterbrock
1983), or iii) a bi-polar scattering nebula (see e.g. Stockton, Ridgway \&
Lilly 1994). The presently favored model is that at least some of the central
optical morphology results from scattering of light from a hidden nucleus
located at the radio core. In the following paragraphs we summarize the key
observations of the nucleus and nuclear environment at radio, X-ray, infrared,
optical and ultraviolet wavelengths.

Radio observations with the VLA at 6 cm wavelength with 0\farcs4 resolution
(Perley, Dreher \& Cowan 1984) show a compact radio core with a knotty,
one-sided jet running along position angle (PA) $284\degmark \pm 2\degmark$
from $\sim 1$ -- 25\arcsec. Higher resolution VLBI observations show a ``core''
with a pc-scale jet that is aligned with the kpc-scale jet to within
$4\degmark$ (Carilli, Bartel \& Linfield 1991).

Near-infrared imaging shows an unresolved nucleus, within 1\arcsec\ of the
radio core, and suggested to be obscured by $A_V \simeq 50 \pm 30$ magnitudes
of extinction (Djorgovski  et al. 1991). Near-infrared spectroscopy by Ward et
al. (1991) detected emission in narrow molecular hydrogen lines, Pa$\alpha$ and
Br$\gamma$. Using an empirical relationship between broad H$\alpha$ flux and
X-ray luminosity for quasars with little reddening, and an observed upper limit
to the flux of broad Pa$\alpha$, these authors estimate an extinction $A_V
({\rm BLR}) > 24$ mag to the broad line region. A similar relationship between
X-ray luminosity and near infrared luminosity, along with an assumption about
the spectral slope of the nuclear spectrum, implies an extinction of $A_V ({\rm
nuc}) \simeq 54 \pm 9$ mag to the nuclear continuum source (Ward et al. 1991).
A more recent study suggests $A_V \sim 150$ -- 200 mag to the nucleus (Ward
1996). For a standard dust to gas ratio $A_V ({\rm mag}) = 5 \times 10^{-22}
N_H (\pcmsq)$, this value of $A_V$ implies an obscuring column density of $N_H
\simeq 3 \times 10^{23} \pcmsq$ to the nucleus.

Within the central few arc-seconds of Cygnus A, there is an optical source with
a bi-polar morphology, the northwest component being dominated by line emission and
the southeast component being dominated by continuum emission (see, e.g. Stockton et
al. 1994). In an HST V band image (Jackson et al. 1996) the central component
resembles the apex of a conical structure similar to that seen in other
galaxies and attributed to light from a hidden nucleus escaping along a
restricted range of solid angle, i.e. an ``ionization cone''.

Ultraviolet observations with the HST FOS by Antonucci, Hurt \& Kinney (1994)
show noisy, broad [\ion{Mg}{2}] with a FWHM $\sim 7500 \kmps$, which is
suggestive of a hidden broad line region. Subsequent optical spectroscopic
observations using Keck II (Ogle et al. 1997) show polarized continuum and
broad Balmer line emission, with the continuum polarization in the northwest
component rising to $\sim$ 16\% at 3800\AA. The polarized light comes from a
bi-conical structure with an opening half-angle in the range $46\degmark$ --
$55\degmark$, corresponding to the ``ionization cone'' seen with HST. The
symmetry of the polarization vectors indicates that this light originated in
the nucleus and has been scattered into our line of sight.

Cygnus A is surrounded by cluster gas and this makes a determination of the
X-ray spectrum of the nucleus difficult. The findings of previous X-ray studies
of the nucleus and cluster gas of Cygnus A are summarized in
Table~\ref{tab:previous_xray}.  The detection of a hard power-law is compelling
evidence for a hidden active nucleus or quasar in the nucleus of Cygnus A and
the values found for the intrinsic column density are typical of Seyfert 2
galaxies (Awaki et al. 1991; Smith \& Done 1996; Turner et al. 1997a). Also,
Arnaud (1996) showed that the hard X-rays come from a point source within
30\arcsec\ of the radio core.

In this paper, we focus on \emph{Chandra} X-ray observations of the nucleus of
Cygnus A. Our results on the X-ray emission from the radio hot spots and the
intra-cluster gas are discussed by Wilson et al. (2000; hereafter Paper I) and
Smith et al. (2001; hereafter Paper III), respectively, while a future paper
will deal with the cavity inflated by the radio jets. A Hubble constant of $H_0
= 50$ km s$^{-1}$ Mpc$^{-1}$ and a deceleration parameter of $q_0 = 0$ have
been assumed throughout. The luminosity and ``angular size'' distances to
Cygnus A are then 346.4 Mpc and 310.6 Mpc, respectively, and 1\arcsec\
corresponds to 1.51 kpc.

\section{Observations and Reduction}

\subsection{Observations}

The nucleus of Cygnus A is known to be a strong X-ray source, and we were
concerned that ``pile-up'' would affect the \emph{Chandra} observations. In
order to obtain the count rate of the nucleus and circumnuclear regions, we
first obtained a short observation (obs id 359) of Cygnus A on 2000 March 8
using chip S3 (backside illuminated) of the Advanced CCD Imaging Spectrometer
(ACIS; Garmire et al. 2000). Single exposures with a 0.1~s frame-time were
alternated with two exposures with a 0.4~s frame-time, the total integration
time being $\la 1$ ks. This observation showed the nucleus was not
significantly piled-up in either the 0.1~s or 0.4~s frame-time, but the
observed count-rate indicated that a 3.2~s frame-time observation of the
nucleus would suffer from significant pile-up. On this basis, we decided to
observe Cygnus A for separate, longer exposures with frame-times of 0.4~s and
3.2~s, which were designed to study primarily the X-ray emission of the nucleus
and the larger scale regions, respectively. The resulting level 1 and level 2
peak count rates in the long exposure 3.2~s frame-time observation (see below)
are 0.0425 and 0.0286 cts pixel$^{-1}$ s$^{-1}$ respectively. Thus only 67\% of
the events were good grades, a clear indication of pile-up. In contrast, the
level 2 and level 1 peak count rates in the long 0.4~s frame-time exposure are
0.053 and 0.056 cts pixel$^{-1}$ s$^{-1}$, respectively, indicating about 94\%
good grades. The higher count rate in the 0.4~s than the 3.2~s frame-time is
also an indicator of pile-up in the 3.2~s frame-time observation.

These longer integrations were taken with the ACIS instrument in two separate
observations. On 2000 May 21 (obs id 360) the standard 3.2~s frame-time was
used. CCDs I2, I3, S1, S2, S3 and S4 were read out, though most of the X-ray
emission from Cygnus A is on S3. On 2000 May 26 (obs id 1707) a sub-array
window 128 pixels wide on the S3 chip was used with a 0.4~s frame-time to
mitigate the effects of pile-up in the inner regions, as discussed above. Both
exposures were taken with Cygnus A at the aim-point of the S3 chip. The data
were screened in the usual way for times of high background count rates and
aspect errors. The 3.2~s and 0.4~s frame-time data have total good exposure
times of 34720~s and 9227~s respectively.  Spectra and instrument responses
were generated using {\sc ciao} 2.1 (with reprocessed data) and CALDB 2.3, and
analyzed with {\sc xspec} v11.0.1. The \emph{Chandra} response is not reliable
below 0.45 keV, although the backside-illuminated S3 chip has good sensitivity
to X-rays down to 0.1 keV. To improve the sampling of the X-ray images, we
interpolated the data from the original $0\farcs5$ to $0\farcs05$-sized pixels
and then smoothed the resulting image with a Gaussian of FWHM $0\farcs5$.

Cygnus A was also observed using the \emph{Rossi X-ray Timing Explorer}
(\emph{RXTE}) on  May 20, 2000. Data were gathered using the Proportional
Counter Array (PCA) and High Energy X-ray Timing Experiment (HEXTE). Two PCA
counters (PCU0 and PCU2) were operating at the time of the observation, but
PCU0 had suffered the loss of its propane layer one week earlier. Because a
good background subtraction model is not yet available  for PCU0 in this new
state we have restricted our analysis to PCU2. We used the standard Rex
script\footnote{\url{http://RXTE.gsfc.nasa.gov/docs/xte/recipes/rex.html}} to
generate PCA and HEXTE light curves, source and background spectra, and
response matrices. Both the PCA and HEXTE have a field of view of $1\degmark$
(FWHM) and are sensitive to photons in the ranges 2 -- 60 keV and 15 -- 250 keV
respectively. The PCA and HEXTE have a spectral resolution of $\Delta E / E
\simeq 18\%$ at 6 keV and 60 keV respectively. The good exposure times were
29.3 ks for the PCA and 9.7 ks for each of the two HEXTE clusters.

\subsection{Extraction Regions}

To obtain a \emph{Chandra} spectrum of the nucleus, counts were extracted from
a 2\farcs5 diameter circle centered on the peak of the X-ray emission. In the
0.4~s frame-time data there are 2781 counts corresponding to a count rate of
$0.301 \pm 0.006$ ct \ps. Background counts were extracted from a concentric
annulus with an inner diameter 7\arcsec\ and outer diameter 20\arcsec. The
background level is very low and is equivalent to only 43 counts, 1.5\% of the
total within the 2\farcs5 diameter circle used for the nucleus.

\label{sec:nw_se_nuc}

In order to investigate the spectrum of the circumnuclear extended regions, we
took counts from two sectors of a circle with diameter $7\arcsec$ centered on
the nucleus. The sectors have half-angles of $55\degmark$ and their axes are
aligned with the jet and counter jet. In the 0.4~s frame-time data there are
3183 counts, corresponding to a count rate of $0.345 \pm 0.006$ cts s$^{-1}$,
within these sectors. Background counts were extracted from a concentric
annulus with inner diameter $7\arcsec$ and outer diameter $14\arcsec$. The
background level is equivalent to 224 counts, 8\% of the total inside the two
sectors. The source extraction regions contain some flux from the nucleus.

\section{X-ray Morphology}

\begin{figure}
  \centerline{
    \includegraphics[scale=0.7]{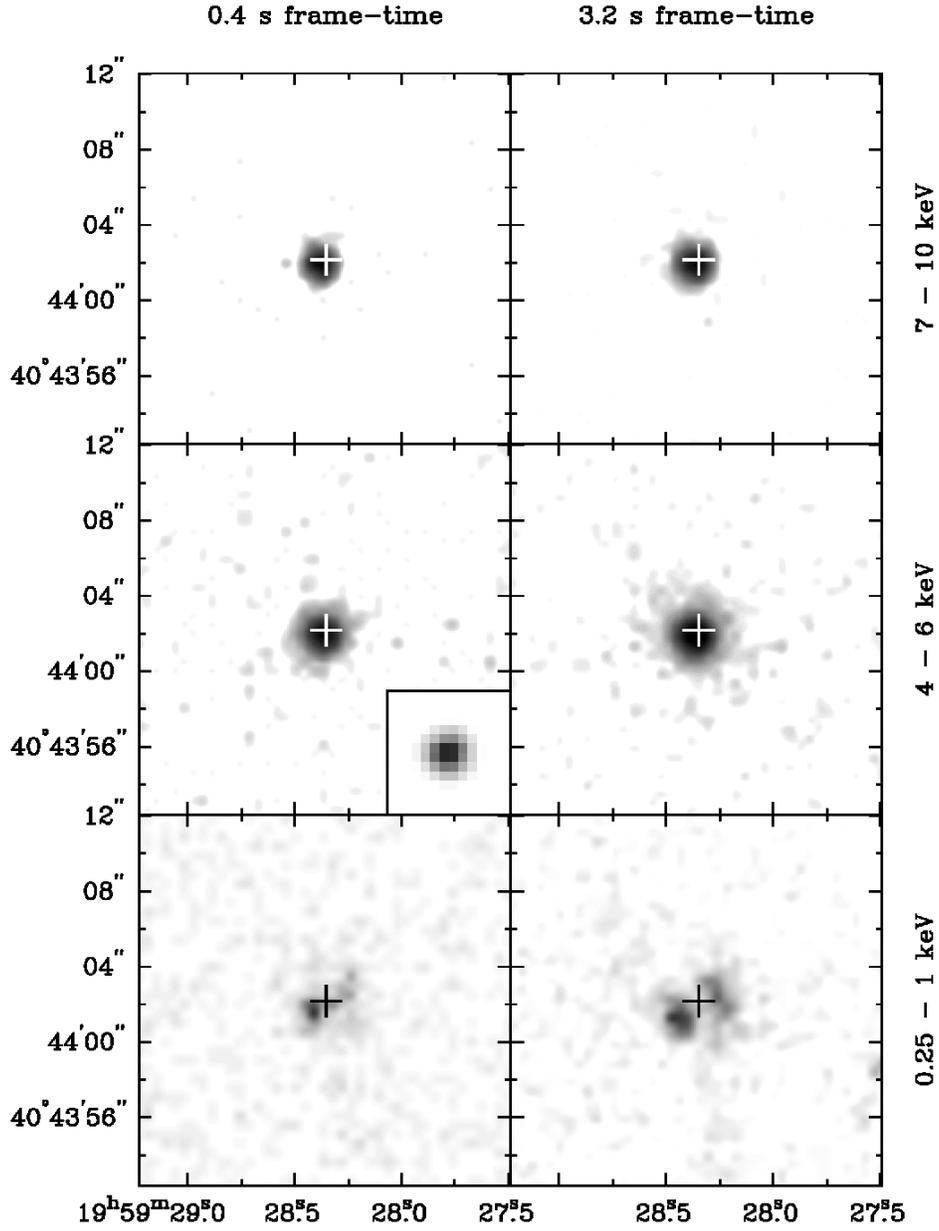}}

\caption{ \label{fig:collage} Grey scale representations of \emph{Chandra}
  X-ray images of the nucleus of Cygnus A in three energy bands and for each of
  the two frame-times used. The \emph{Chandra} images have been resampled to
  ten times smaller pixel size and smoothed by a Gaussian of FWHM $0\farcs5$.
  The left column has a 0.4~s frame-time and the right column a 3.2 ~s
  frame-time. In the top row the energy range is 7 -- 10 keV, the center row 4
  -- 6 keV and the bottom row 0.25 -- 1 keV. The center left panel also shows
  the point spread function at 4.51 keV. The `+' symbol indicates the location
  of the peak of the hard X-ray emission that has been aligned with the radio
  core. The lookup tables for the grey scale are different in each panel, with
  the upper two rows having a logarithmic scale and the bottom row a linear
  scale.}

\end{figure}

\begin{figure}
  \centerline{
    \includegraphics[angle=0,scale=0.8]{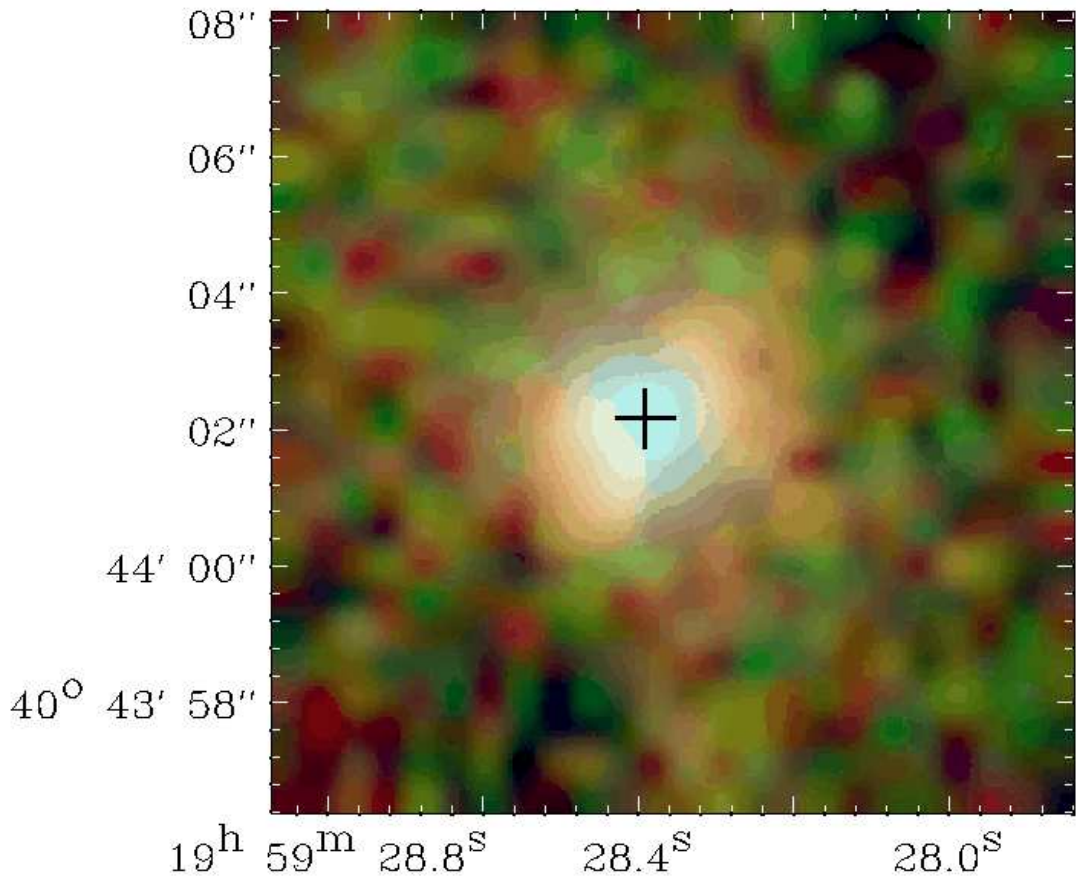}}

\caption{ \label{fig:color} Color representation of the \emph{Chandra} X-ray
  image of the nucleus of Cygnus A. The image has been resampled to ten times
  smaller pixel size and smoothed by a Gaussian of FWHM $0\farcs5$. The data
  were divided into three bands, red representing 0.1 -- 1.275 keV, green 1.275
  -- 2.2 keV and blue 2.2 -- 10 keV. The nuclear region has a blue (hard)
  point-like core with two offset regions of softer emission. This emission is
  embedded in a comparatively smooth distribution of hot gas. The `+' symbol
  indicates the location of the peak of the hard X-ray emission that has been
  aligned with the radio core.}

\end{figure}

The X-ray morphology within the central $4 \arcsec \times 4 \arcsec$ around the
nucleus varies strongly with photon energy (see Fig.~\ref{fig:collage}). Above
approximately 2 \keV\, the nucleus appears to be point-like and coincident with
the radio core (see Section~\ref{sec:x_ray_align}), while below $\sim 1 \keV$ a
bi-polar distribution is clearly seen in the higher signal-to-noise ratio 3.2~s
frame-time data, with one peak of emission $\simeq 1\farcs2$ northwest of the
nucleus along PA ${287^{+35\degmark}_{-13\degmark}}$ and a second peak $\simeq
1\farcs3$ southeast of the nucleus along PA
${115^{+30\degmark}_{-21\degmark}}$. These are the position angles of the
center of the brightest pixels, in the 0.25 -- 1 keV image, to the northwest
and southeast relative to the location of the nucleus and determined using the
{\sc iraf} {\sc center} tool. The quoted ``error bars'' are the range of
position angle given by moving $\sqrt{2}$ pixels from the peak pixel diagonally
NE and SW. The bi-polar distribution is easily seen in a color image
(Fig.~\ref{fig:color}), in which the data were divided into three bands, red
representing 0.1 -- 1.275 keV, green 1.275 -- 2.2 keV and blue 2.2 -- 10 keV.
Images in the three bands were then combined to give the overall color picture.
The nuclear region has a blue, and hence hard, X-ray point-like core with the
two offset regions to the northwest and southeast appearing yellower,
indicative of softer X-ray emission. Even though a clear bi-polar morphology is
seen, there is also soft X-ray emission close to the nucleus. A comparatively
smooth distribution of hot gas is projected around the nuclear region.

The double morphology may be due, in part, to the presence of a dust lane
running N-S across the nucleus obscuring the central $\sim 1\arcsec$ of the
soft X-ray emission. The 0.25 -- 1 keV flux per pixel at the location of the
nucleus is approximately 40\% of that at the peak of the soft X-ray emission
some $\sim 1\farcs2$ from the nucleus. If the intrinsic luminosity and spectra
of these two regions are the same then a modest additional column density of
$N_H ({\rm cold\ gas\ in\ dust\ lane}) \simeq 2.4 \times 10^{21} \pcmsq$
obscuring the nucleus would be sufficient to account for the lower flux at the
nucleus. This column density is consistent with the additional reddening of the
nucleus by the dust lane (Shaw \& Tadhunter 1994; Tadhunter, Metz \& Robinson
1994), assuming a standard dust to gas ratio.

\begin{figure}
  \includegraphics[angle=270, scale=0.65]{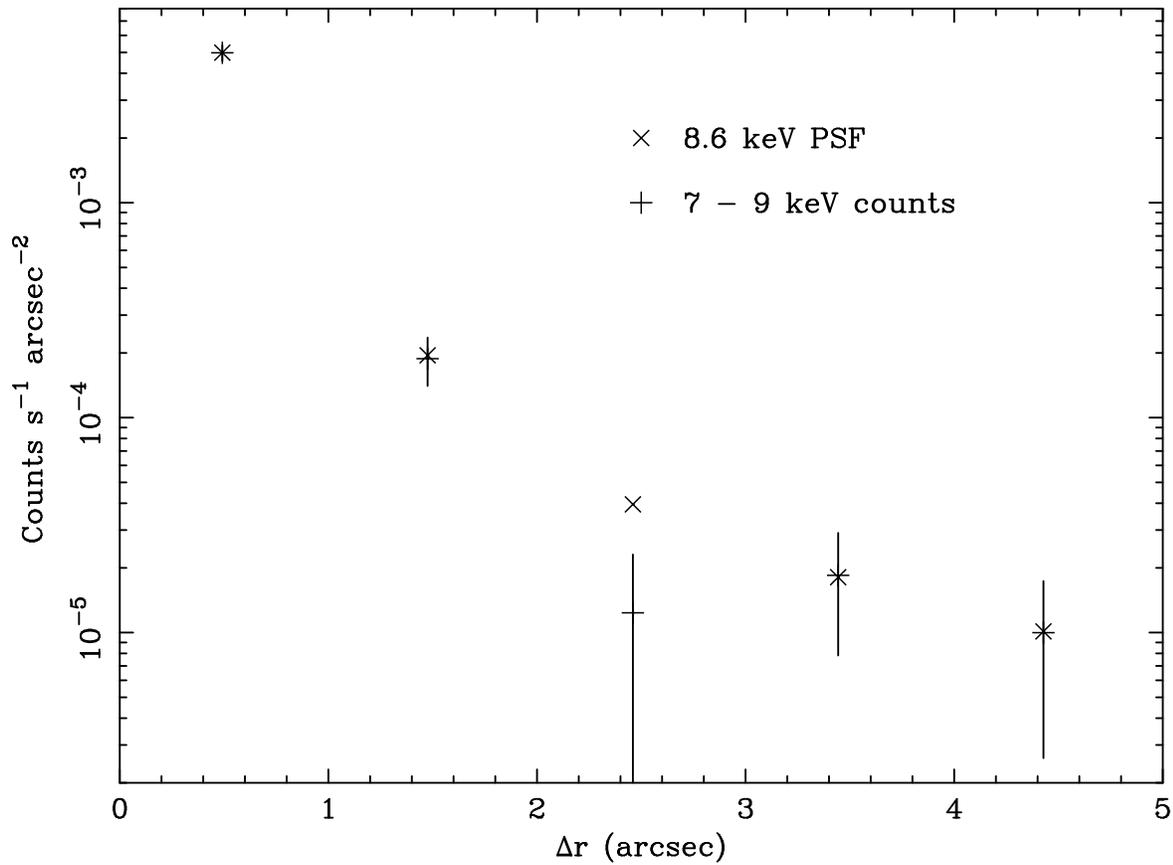} 

\caption{ \label{fig:psf} The radial profile of counts from the nucleus of
  Cygnus A extracted from the 0.4~s frame-time data in the energy range 7 -- 9
  keV (`$+$' points with Poissonian error bars) compared to the theoretically
  computed PSF at 8.6 keV (`$\times$' points). The data indicate the hard X-ray
  source in the nucleus is unresolved.}

\end{figure}

To determine whether \emph{Chandra} resolves the point-like, hard X-ray
nucleus, its radial profile was compared with a model of the point spread
function. It should be noted that on larger spatial scales, significant X-ray
emission is observed in the neighborhood of the nucleus. Most of this larger
scale emission is thermal in origin with a temperature of 4 -- 5 keV and
apparently represents emission from the intracluster gas projected close to the
nucleus (Paper III). This hot plasma has a significant high energy tail so the
wings of the PSF from the nucleus may merge into the continuum emission from
the thermal plasma. To minimize this effect, the point spread function was
calculated at 8.6 keV where the contribution from the cluster gas is much
smaller than at lower energies. The background level was determined using a
circle of radius 22\farcs5 offset 50\arcsec\ southeast of the nucleus. Radial profiles
of the point spread function at 8.6 keV and the observed 7 -- 9 keV counts
(from the 0.4 s frame-time data) centered on the nucleus are shown in
Fig.~\ref{fig:psf}. The hard X-ray nucleus is seen to be  unresolved.

\section{X-ray Spectra}

\subsection{Nucleus} \label{sec:nuc_spec}

\begin{figure}
  \centerline{
    \includegraphics[angle=270, scale=0.65]{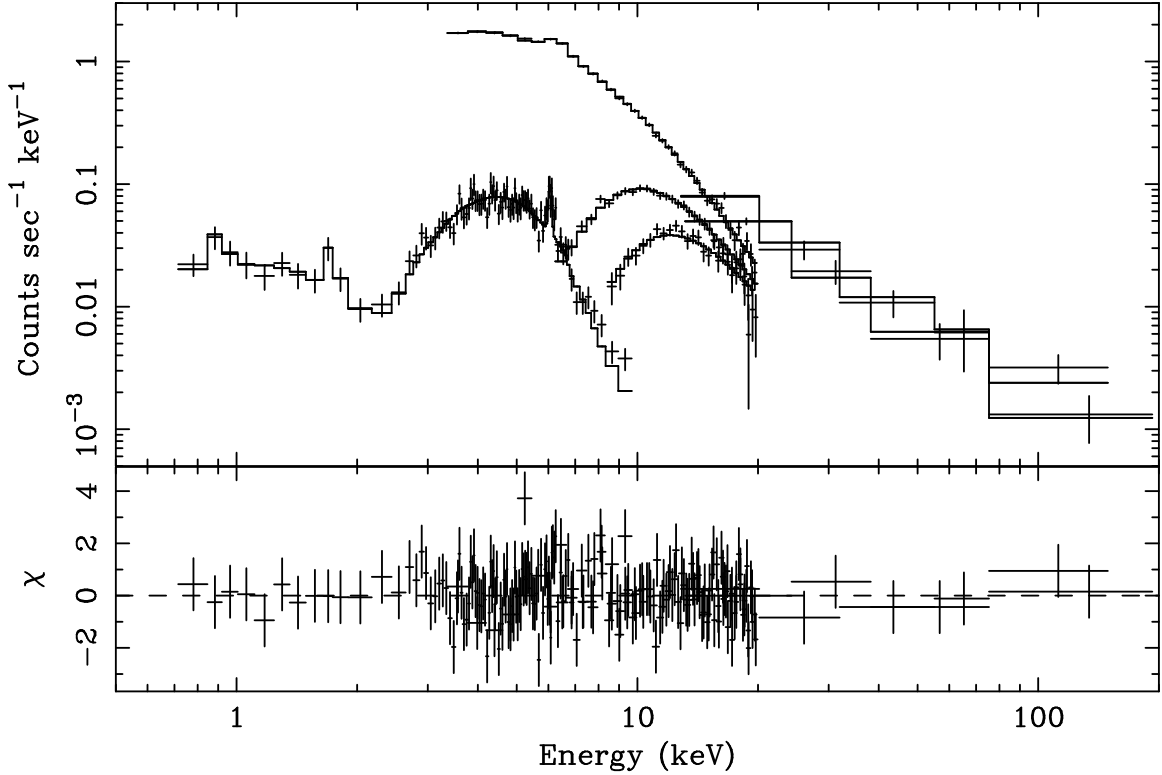}}

\caption{ \label{fig:spec} X-ray spectra of Cygnus A extracted from the 0.4 s
  frame-time \emph{Chandra} data and the \emph{RXTE} data. The energy scale is
  as observed. The upper panel shows the data points with error bars (crosses),
  with the model folded through the instrument responses (solid lines passing
  through the data points). The lower panel shows the $\chi$ residuals from
  this fit. The parameters of the model (``Model 1'') are listed in Table
  \ref{tab:nuc_spec}. The spectrum over the range 0.7 -- 9 keV is the
  \emph{Chandra} observation of the nucleus through a circular aperture of
  diameter 2\farcs5 with a 0.4~s frame-time. The curve near the top of the
  figure at 3 keV and extending from 3 to 20 keV is the \emph{RXTE} layer 1 PCA
  spectrum, that extending from 6 to 20 keV is the layer 2 PCA spectrum, and
  that extending from 8 to 20 keV is the layer 3 PCA spectrum. The HEXTE
  spectrum extends from 12 to 200 keV.}

\end{figure}

The grossly different spatial resolutions of \emph{Chandra} and \emph{RXTE}
mean that different spectral models have to be fitted to data sets from each
observatory. The 2\farcs5 diameter circle used to extract counts from the
\emph{Chandra} data effectively isolates the nucleus from the surrounding
cluster, whereas the $\simeq 1\degmark$ resolution of \emph{RXTE} yields a
spectrum that includes the nucleus, features associated with the radio source,
and the intra-cluster gas. The \emph{Chandra} data show the ``hot spots'' to
have $\approxlt 1\%$ of the 7 -- 10 keV flux of the nucleus so flux from the
``hot spots'' will not contribute significantly to the \emph{RXTE} spectrum.
The \emph{Chandra} and \emph{RXTE} spectra are shown in Fig.~\ref{fig:spec}.
The \emph{Chandra} spectrum of the nucleus has two main components --- a soft
component below 2 keV, and a hard component above 2 keV. In the previous
Section, we noted that the soft component is spatially extended and the hard
component spatially unresolved. Additionally there is a strong iron line around
6.4 keV in the rest frame of Cygnus A. The \emph{Chandra} spectrum was modeled
over the energy range 0.7 -- 9 keV. The \emph{RXTE} spectra cover a much larger
spatial area and extend up to much higher energies. The PCA spectra were fitted
over energy ranges of 3 -- 20 keV (layer 1), 6 -- 20 keV (layer 2) and 8 -- 20
keV (layer 3). The HEXTE spectra were fitted over the energy range 15 -- 200
keV, and grouped (using the {\sc ftool} {\sc grppha}) into bins such that the
signal to noise ratio in each bin exceeded 40 (i.e. $(s_i - b_i) / \sqrt{s_i +
b_i} > 40$ where $s_i$ are the measured source counts and $b_i$ are the
background counts in bin $i$). The various spectral components were modeled as
follows, i) a heavily absorbed power law (with photon index $\Gamma_h$, where
the ``h'' stands for ``high energy'') to represent the hidden quasar, ii)
either a power law (with photon index $\Gamma_l$, where ``l'' stands for ``low
energy'') plus narrow emission lines (``Model 1'') or a bremsstrahlung plus
narrow emission lines (``Model 2''), all absorbed by the Galactic column
density (assumed here to be $N_H ({\rm Gal}) = 3.3 \times 10^{21} \pcmsq$
[Dickey \& Lockman 1990]), to describe the extended soft X-ray emission (there
is only very marginal evidence for intrinsic absorption of this emission -- see
Section \ref{sec:column}), iii) a ``neutral'' iron fluorescence line, and iv) a
high temperature bremsstrahlung with two narrow emission lines to represent the
thermal emission of the intra-cluster gas. Components i), ii) and iii) were
fitted to all of the data sets, while component iv) was fitted to only the
\emph{RXTE} data sets.  The six data sets -- one \emph{Chandra}, three
\emph{RXTE} PCA and two \emph{RXTE} HEXTE -- were modeled simultaneously using
{\sc xspec}, with the relative normalizations of the \emph{Chandra}, PCA and
HEXTE data allowed to vary. The parameters of these models are listed in Table
\ref{tab:nuc_spec}. All quantities refer to the rest-frame of Cygnus A, with
the exception of the column density to the low energy nuclear emission, which
refers to the observer's frame. The relative normalizations in
Table~\ref{tab:nuc_spec} give the factors by which the calibrated spectra from
the various instruments have been multiplied to provide the best fit. The
factor for \emph{Chandra} is defined to be 1.0 and those for the \emph{RXTE}
detectors are close to or consistent with unity.

\begin{figure}
  \includegraphics[angle=270, scale=0.65]{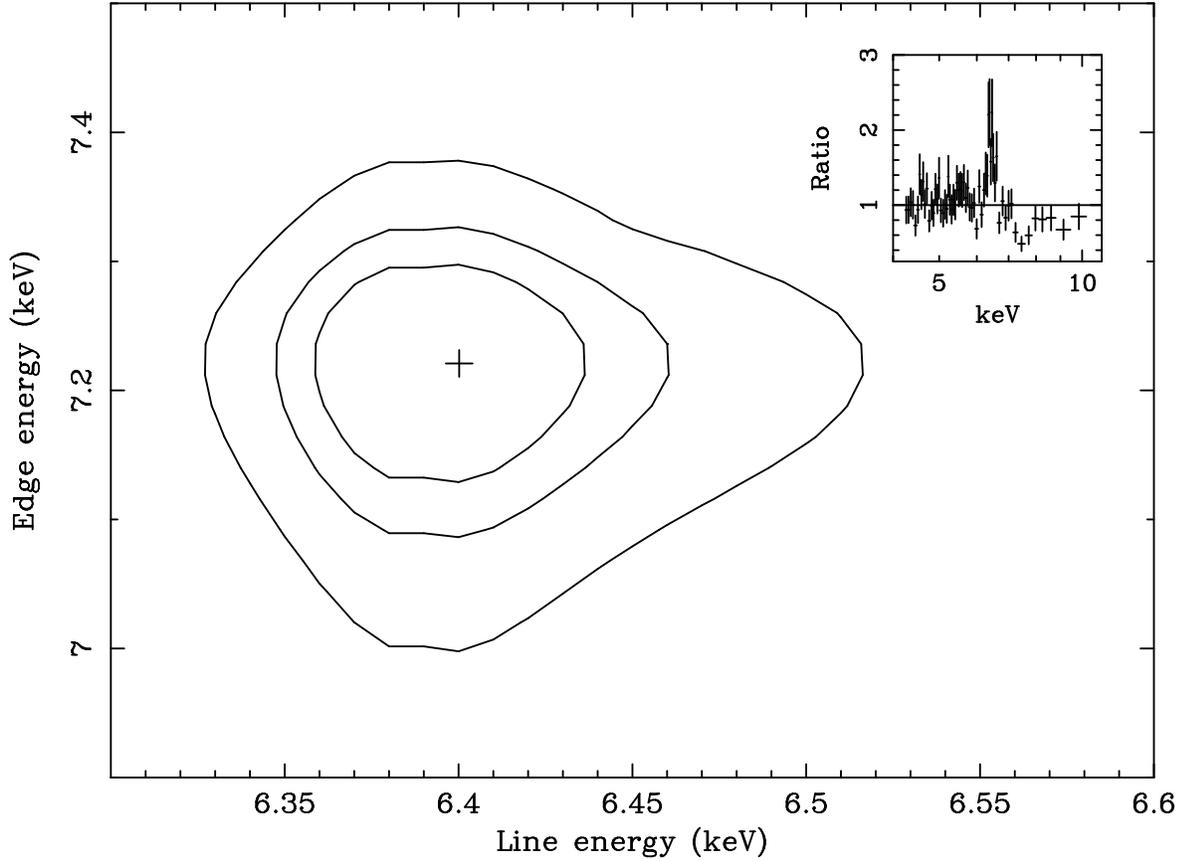}
  
\caption{ \label{fig:fe_line_edge} Confidence contours (plotted at 68\%, 90\%
  and 99\%) of the rest-frame energy of the iron line and absorption edge from
  modeling the 0.4~s frame-time \emph{Chandra} data. The aperture is a circle
  of diameter $2\farcs5$. The line energy is consistent with that of the Fe
  K$\alpha$ fluorescence line of neutral iron. The absorption edge is
  consistent with that of iron in the range \ion{Fe}{1} -- \ion{Fe}{10}. The
  panel in the upper right show the ratio of the \emph{Chandra} data to a power
  law as a function of rest-frame energy. To produce this panel, the data were
  modeled by a power law from 4 to 9 keV.}

\end{figure}

In ``Model 1'' the low energy ($<2 \keV$) part of the nuclear spectrum is
modeled with a power law, the photon index ($\Gamma_l$) of which is constrained
to be the same as that ($\Gamma_h$) of the high energy power law describing the
hidden nucleus. Such would be the case if the low energy emission represents
electron-scattered light from the hidden nucleus. This model provides an
excellent description of the data with a \chisq\ of 181 for 185 degrees of
freedom (d.o.f.). In ``Model 2'' the low energy part of the spectrum is modeled
as bremsstrahlung emission. This also provides an excellent description of the
data with a \chisq\ of 177 for 184 d.o.f. (Table~\ref{tab:nuc_spec}). In both
models the hidden nucleus is described by a power law, the best fit value of
which has $\Gamma_{h1} (= \Gamma_{l1}) = 1.52^{+0.12}_{-0.12}$ (the subscript 1
denotes ``Model 1''), $\Gamma_{h2} = 1.53^{+0.12}_{-0.15}$ (the subscript 2
denotes ``Model 2''), obscured by a column density of $2.0^{+0.1}_{-0.2} \times
10^{23} \pcmsq$. These values of $\Gamma$ are consistent with that obtained --
$\Gamma_l = 1.80^{+1.1}_{-1.1}$ -- by simply modeling the soft (0.6 -- 2.5 keV)
spectrum by an absorbed power-law. The HEXTE data alone are well described by a
power law of photon index $\Gamma_h = 1.48^{+0.36}_{-0.34}$, between 20 and 200
keV. These values for $\Gamma_h$ are similar to those found for unobscured AGN,
such as Seyfert 1 galaxies. The unabsorbed rest-frame 2 -- 10 keV luminosity of
the hidden nucleus in Cygnus A is $L_X ({\rm nuc}) = 3.7 \times 10^{44}
\ergps$, and its 0.5 -- 2 keV luminosity $1.5 \times 10^{44} \ergps$. In Model
1, the soft X-ray component is absorbed by a column density of
$1.9^{+1.7}_{-1.0} \times 10^{21} \pcmsq$ which is nominally lower than, but
consistent with, the Galactic value. The \emph{Chandra} data show soft X-ray
lines from He-like and/or H-like Ne and ionized Si from the nucleus (see Table
\ref{tab:nuc_lines}). A strong, narrow nuclear Fe K$\alpha$ fluorescence line
is seen at $6.40^{+0.05}_{-0.03}$ keV, consistent with \ion{Fe}{2} --
\ion{Fe}{18}. If the line width is a free parameter we obtain $\sigma_{\rm Fe}
= 0.09^{+0.03}_{-0.04}$ keV which is not significantly broad, and comparable to
the detector resolution at 6 keV, $\sigma_{\rm detector} \simeq 0.08$
keV\footnote{In the present calibration the FWHM of the line response is
systematically narrower than the data, so $\sigma_{\rm Fe}$ is consistent with
zero.}. The equivalent width (EW) of the Fe line, $182^{+40}_{-54}$ eV, is in
excellent agreement with theoretical predictions for the EW versus $N_H ({\rm
nuc})$ relation in various geometries (see Fig. 3 of Ptak et al. 1996). A
strong, nuclear, iron absorption edge is observed at a rest-frame energy of
approximately 7.2 keV, corresponding to the K shell edge of \ion{Fe}{1} --
\ion{Fe}{10} (see Fig.~\ref{fig:fe_line_edge}). If the iron abundance in the
absorbing gas is allowed to vary (using the {\sc vphabs} model in {\sc xspec})
the best fitting value is found to be $Z ({\rm Fe}) = 0.90^{+0.46}_{-0.48}
\times Z_\odot ({\rm Fe})$. In addition, there is tentative evidence for an
iron K$\beta$ line at a rest-frame energy of 7.1 keV, and this will add to the
uncertainty in the estimate of the iron abundance.

The temperature of the thermal emission from the cluster gas, measured by the
\emph{RXTE} instruments, is $kT_1 = 6.9^{+0.3}_{-1.0} \keV$, $kT_2 =
6.8^{+0.5}_{-0.8} \keV$, and the unabsorbed rest-frame 2 -- 10 keV luminosity
of the cluster is $L_X ({\rm cluster}) = 1.2 \times 10^{45} \ergps$. These
parameters for the cluster are in good agreement with  the results from
\emph{Ginga} (Ueno et al. 1994), \emph{EXOSAT} (Arnaud et al. 1987) and
\emph{ASCA} (Sambruna et al. 1999; Markevitch et al. 1998, 1999). The cluster
flux we measure with \emph{RXTE} is larger than that found from modeling the
\emph{Chandra}-detected emission in chip S3 (an $8^\prime \times 8^\prime$
region of the inner part of the cluster gas --- see paper III) since the field
of view of \emph{RXTE} is much larger than that of \emph{Chandra}. The model of
the cluster gas includes two strong emission lines, one at a rest-frame energy
of 6.70 keV from ionized Fe K$\alpha$, and another at 7.90 keV (see Table
\ref{tab:nuc_lines}) that may be either or both of ionized Fe K$\beta$ and
ionized Ni K$\alpha$. 

\subsection{Northwest and Southeast Regions} \label{sec:nw_se_spec}

In order to study the extended soft X-ray emission within a few arc seconds of
the nucleus, we obtained a spectrum from the 0.4~s frame-time data using two
sectors of a circle with diameter $7\arcsec$ centered on the nucleus. The
sectors had half-angles of $55\degmark$ and their axes were aligned with the
jet and counter jet (cf. Fig.~\ref{fig:collage}, bottom panels). We restrict
our attention to the energy range 0.5 -- 2 keV to minimize the contribution
from the unresolved nucleus. The spectrum was initially modeled by a hot
thermal plasma with solar metal abundances using the {\sc mekal} model in {\sc
xspec}, absorbed by the Galactic column. This model provides a very poor fit to
the data (see Table~\ref{tab:nw_se_spec}) and may be ruled out. If the metal
abundances are allowed to vary, the best model has an abundance of only 3\% of
solar, which is implausible. Bremsstrahlung or power-law models, both absorbed
by the Galactic column, are also unsatisfactory (Table~\ref{tab:nw_se_spec})

\begin{figure}
  \includegraphics[angle=270, scale=0.65]{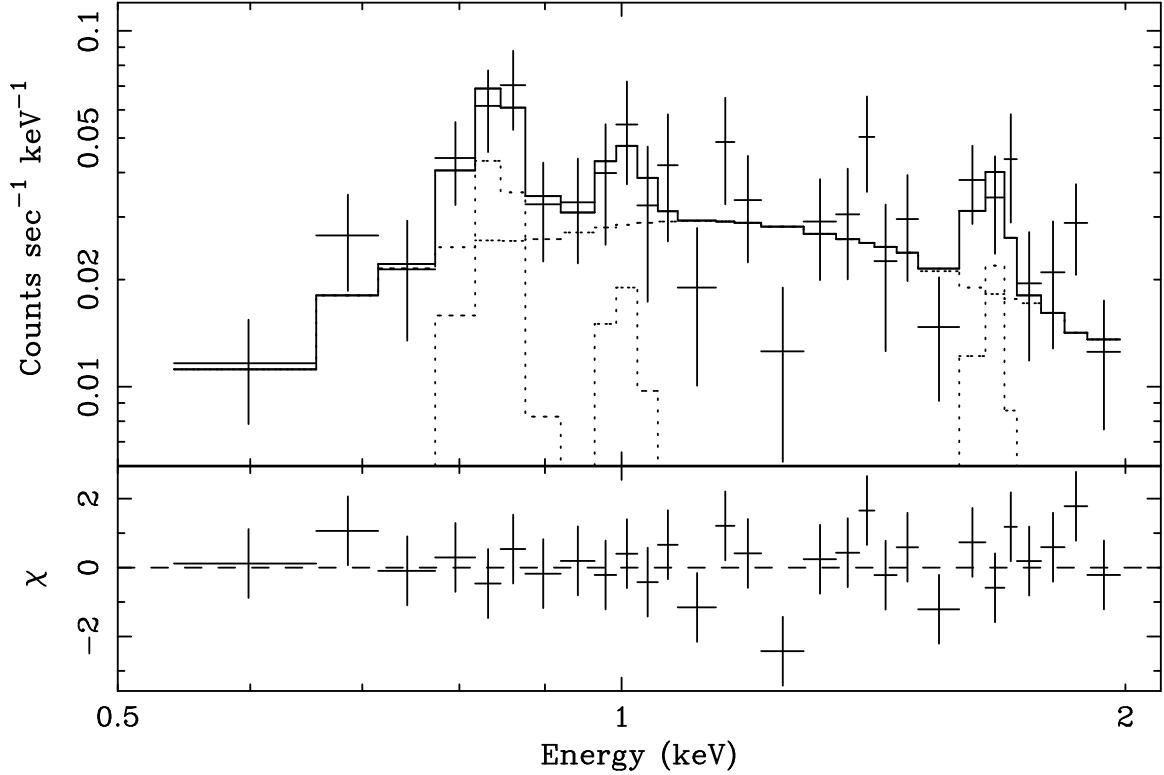}

\caption{ \label{fig:nw_se_spec} X-ray spectrum of the soft extended nuclear
  emission of Cygnus A extracted from the 0.4~s frame-time data
  (Section~\ref{sec:nw_se_spec}). The aperture consisted of two sectors of a
  circle with diameter $7\arcsec$ centered on the nucleus, each sector having
  an opening half-angle of $55\degmark$ and their axes aligned with the jet and
  counter jet. The energy scale is as observed. The upper panel shows the data
  points with error bars (crosses) and the model folded through the instrument
  response (uppermost solid line passing through the data points). The
  individual components of the model are plotted below this with dotted lines.
  The lower panel shows the $\chi$ residuals to this fit. The parameters of
  this model are listed in Tables \ref{tab:nw_se_spec} and
  \ref{tab:nw_se_lines}.}

\end{figure}

A power law plus narrow emission lines, all absorbed by the Galactic column,
provides a much better description of the data (see Tables
\ref{tab:nw_se_spec}, \ref{tab:nw_se_lines} and Fig.~\ref{fig:nw_se_spec}),
with a $\chi^2$ value of 22 for 21 degrees of freedom (d.o.f.). The photon
index of the power law, $\Gamma_l = 2.01^{+0.46}_{-0.54}$, is consistent with
that obtained through the 2\farcs5 aperture, but marginally steeper than that
found for the heavily absorbed nuclear source at high energy, $\Gamma_h =
1.52^{+0.12}_{-0.12}$ (Model 1), or $\Gamma_h = 1.53^{+0.12}_{-0.15}$ (Model 2;
see Table~\ref{tab:nuc_spec}). For the power law plus narrow emission lines
model, the unabsorbed rest-frame 0.5 -- 2 keV luminosity is $3.8 \times 10^{42}
\ergps$ and, by extrapolating to higher energies with $\Gamma_l = 2.01$, the
unabsorbed rest-frame 2 -- 10 keV luminosity is found to be $3.7 \times 10^{42}
\ergps$ (7\arcsec\ diameter, double sector aperture). For comparison, the power
law plus lines model through the 2\farcs5 aperture (Model 1) has a rest-frame
0.5 -- 2 keV luminosity, corrected for Galactic absorption, of $2.0 \times
10^{42} \ergps$ and, by extrapolating the low energy power law to higher
energies with $\Gamma_l = 1.52$, an unabsorbed rest-frame 2 -- 10 keV
luminosity of $4.2 \times 10^{42} \ergps$.

We observe lines from \ion{Ne}{7} -- \ion{Ne}{9} K$\alpha$ and \ion{Si}{2} --
\ion{Si}{11} K$\alpha$, and tentatively detect a H-like Ne K$\alpha$ line (see
Table \ref{tab:nw_se_lines}). The spectrum is shown in
Fig.~\ref{fig:nw_se_spec}. If we fit a power law from 4 -- 9 keV plus a narrow
Gaussian we find an iron line consistent with \ion{Fe}{2} -- \ion{Fe}{18}
K$\alpha$ with an equivalent width of 226~eV. In the 3.2~s frame-time data we
observe two lines, \ion{Ne}{5} -- \ion{Ne}{9} K$\alpha$ and \ion{Si}{7} --
\ion{Si}{10} K$\alpha$, consistent with the 0.4~s frame-time data, but do not
require the H-like Ne line. The line fluxes from the 3.2~s frame-time
observation are too low because of the effects of pile-up, and we do not give
them.

\begin{figure}
  \includegraphics[angle=270, scale=0.65]{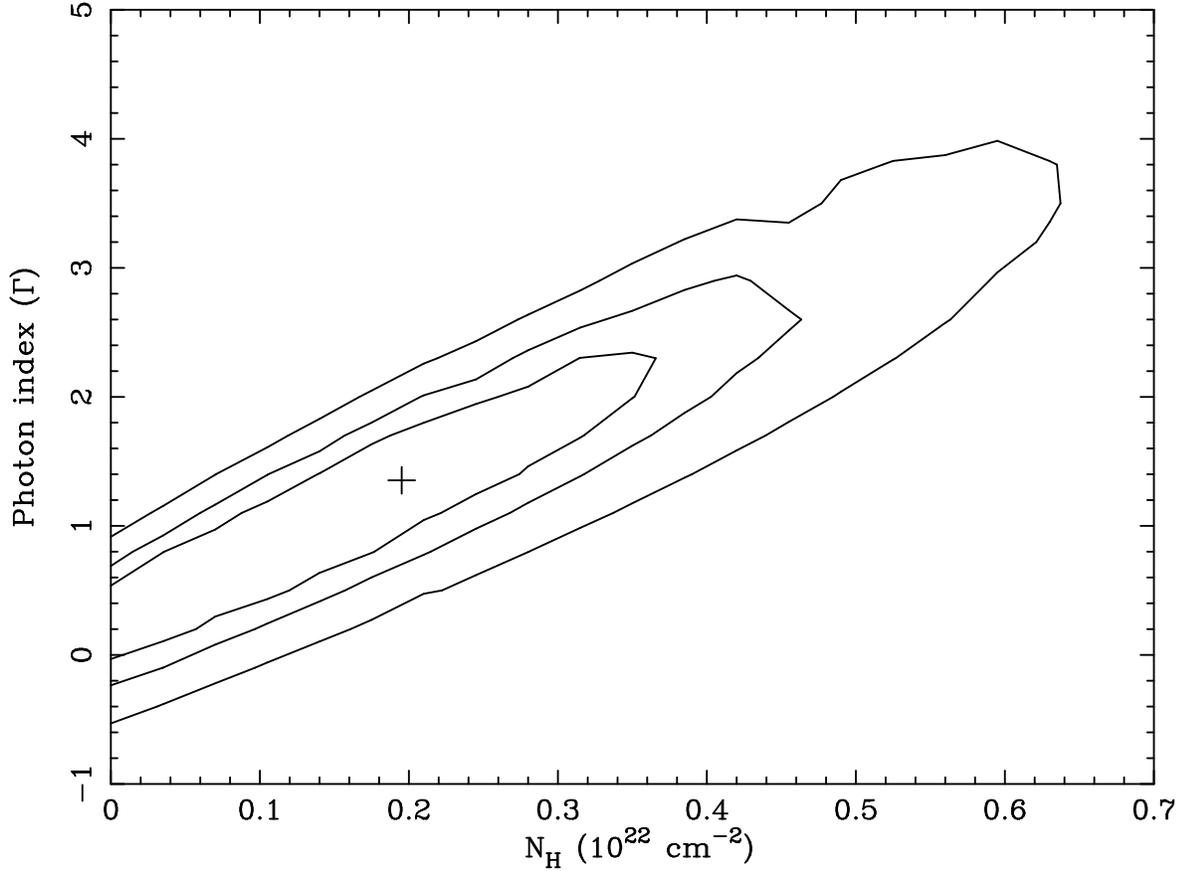}
  
\caption{ \label{fig:gamma_nh} Confidence contours of the absorbing column
density $N_H$ and photon index $\Gamma_l$ of the soft X-ray emission plotted at
68\%, 90\% and 99\%. The aperture consists of two sectors of a circle with
diameter $7\arcsec$ centered on the nucleus, each sector having an opening
half-angle of $55\degmark$ and their axes aligned with the jet and counter jet.
Note that $\Gamma_l$ and $N_H$ are not independent, and an error in the value
of the assumed column density will give a corresponding error in the value of
$\Gamma_l$. The Galactic column density is assumed to be $3.3 \times 10^{21}
\pcmsq$ in this paper.}

\end{figure}

It should be noted that the value of $\Gamma_l$ depends on the assumed
absorbing column density. As can be seen from Fig.~\ref{fig:gamma_nh}, the
lower the absorbing column density the lower the inferred photon index, with an
uncertainty of $\Delta N_H \simeq 1 \times 10^{21} \pcmsq$ corresponding to an
uncertainty in the photon index of $\Delta \Gamma_l \simeq 0.6$. The precise
values of the extinction towards, and intrinsic to, Cygnus A are difficult to
determine, given its low galactic latitude. Within Cygnus A itself the
extinction is very patchy (see, e.g., Fig. 7 of Jackson et al. 1998), while the
foreground reddening, measured towards three normal elliptical galaxies in the
Cygnus A group, is approximately $E(B-V) = 0.36$ (Spinrad \& Stauffer 1982).
Assuming a standard gas to dust ratio this foreground extinction corresponds to
a column density of $2.2 \times 10^{21} \pcmsq$. If the average absorbing
column density over the $7\arcsec$, double sector aperture is close to this
value instead of the assumed $N_H ({\rm Gal}) = 3.3 \times 10^{21} \pcmsq$, the
value of $\Gamma_l$ becomes $\simeq 1.5$, the same value as found for
$\Gamma_h$ (Section \ref{sec:nuc_spec}).

To investigate any spectral differences between the northwest and southeast regions the
circumnuclear emission was divided into two regions. A circle of diameter
10\arcsec\ centered on the peak of the X-ray emission was divided into two
semi-circles by a line running along PA 14$\degmark$ which is perpendicular to
the radio jet on small scales. Also, a square region of $3 \times 3$ pixels
($1\farcs5 \times 1\farcs5$) centered on the peak pixel was excluded. The 0.4 s
frame-time data between 0.5 and 2 keV from each region were modeled by an
absorbed power law of fixed photon index $\Gamma = 2.10$ (cf. Table
\ref{tab:nw_se_spec}). The best model has column densities of $N_H$ (NW) $ =
2.6^{+0.9}_{-0.8} \times 10^{21} \pcmsq$, consistent with the Galactic column
density towards Cygnus A, and $N_H$ (SE) $ = 4.7^{+1.5}_{-1.1} \times 10^{21}
\pcmsq$. This possible excess column to the southeast side suggests it is the farther,
in agreement with Carilli \& Barthel (1996). If real, the difference in
absorbing columns would correspond to $\Delta A_V \simeq 1$ mag of optical
extinction. \label{sec:column}

\section{Comparison with Observations at Other Wavelengths}

\begin{figure}
  \includegraphics[angle=0, scale=0.80]{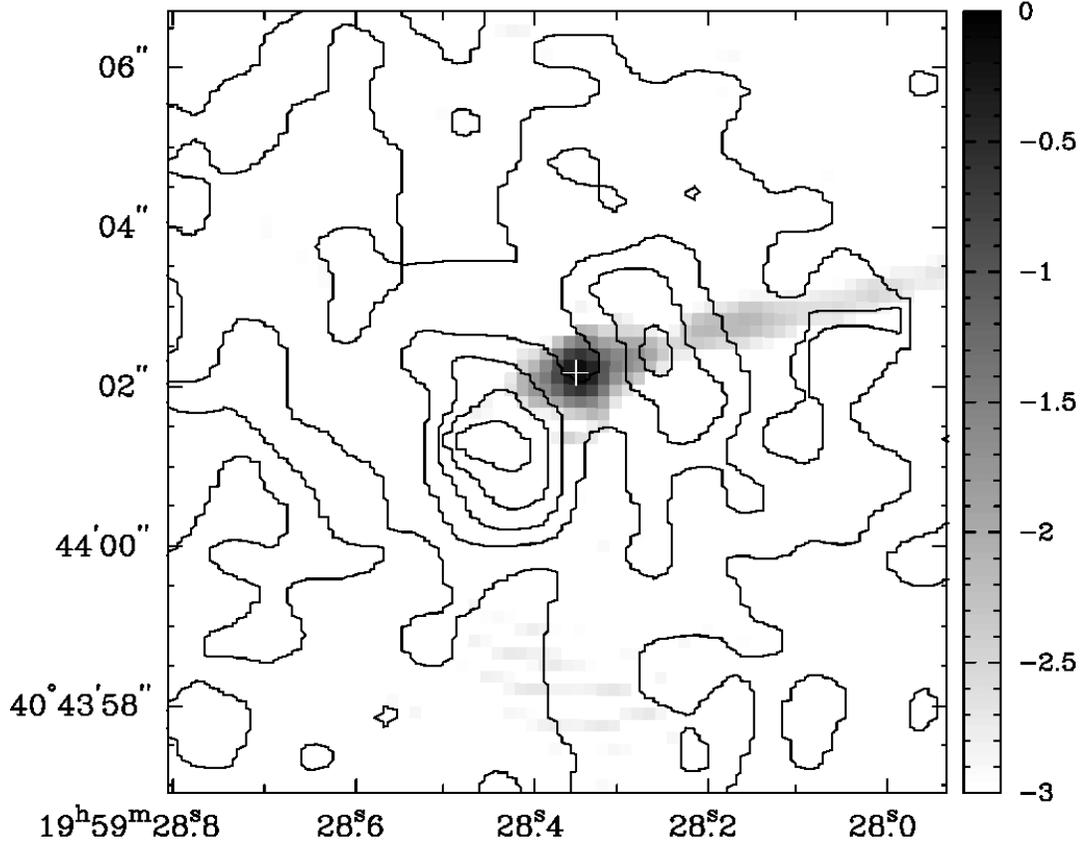}

\caption{ \label{fig:chandra_6cm} A superposition of the \emph{Chandra} 3.2 s
  frame-time data with events in the range 0.25 to 1.00 keV (contours) on a 6
  cm radio map (grey scale; Perley et al. 1984). The resolution (FWHM) of the
  radio map is 0\farcs4. The contours of the \emph{Chandra} image are plotted
  at 2, 6, 10, 14 and 18 cts ($0\farcs5$ pixel)$^{-1}$. The \emph{Chandra}
  image has been resampled to ten times smaller pixel size and then smoothed by
  a Gaussian of FWHM $0\farcs5$. The grey scale of the radio map is
  proportional to the logarithm of the intensity between 0.001 (white) and 1 Jy
  (beam)$^{-1}$ (black), as indicated by the vertical bar. Note the ``neck'' in
  the radio jet 1\farcs2\ to the northwest of the nucleus, where it passes through the
  northwest extended X-ray emission. This ``neck'' also corresponds to a channel
  dividing the northwest cloud seen in HST images in bands containing redshifted
  [\ion{O}{3}] $\lambda 5007$ (see Fig.~\ref{fig:hst_oiii}) and H$\alpha$ (see
  Fig.~\ref{fig:hst_halpha}).}

\end{figure}

The central few arc seconds of Cygnus A contain considerable structure
(Section~\ref{sec:intro}). Radio observations (Perley, Dreher \& Cowan 1984;
Carilli, Bartel \& Linfield 1991) show a compact ``core'' located at
$\alpha_{\rm r} = 19^{\rm h}\ 59^{\rm m}\ 28\fs348$, $\delta_{\rm r} =
40^\circ\ 44^\prime\ 02\farcs17$ (J2000) with a prominent jet running along PA
$284 \pm 3^\circ$ (Carilli et al. 1991).  We assume the core to be the location
of the ``true'' nucleus, i.e.  the putative super-massive black hole and
accretion disk. The peak of the X-ray emission, found using the {\sc iraf} tool
``{\sc center}'' and the \emph{Chandra} astrometry, is at $\alpha_{\rm x} =
19^{\rm h}\ 59^{\rm m}\ 28\fs375$, $\delta_{\rm x} = 40^\circ\ 44^\prime\
02\farcs36$ (J2000). The offset between these two positions is less than
$0\farcs4$ (i.e., smaller than one \emph{Chandra} pixel and within the current
systematics) and we assume them to be coincident. The \emph{Chandra} images
were, therefore, shifted so the peak of the X-ray emission coincides with the
peak of the radio emission. A superposition of the \emph{Chandra} 3.2~s
frame-time data on a 6 cm radio map is shown in Fig.~\ref{fig:chandra_6cm}. The
two peaks of soft X-ray emission, one $\simeq 1\farcs2$ along PA 287$\degmark$
to the northwest and the other $\simeq 1\farcs3$ along PA 115$\degmark$ to the
southeast are, to within a few degrees, along the PA of the radio jet and
counter jet, respectively. The peak to the northwest coincides with a ``neck''
in the radio jet, and also corresponds to a channel dividing the northwest
cloud seen in HST images (discussed below). There is evidence of a counter jet
in the radio observations (Carilli \& Barthel 1996), and this passes just to
the north of the southeast X-ray peak. \label{sec:x_ray_align}

\begin{figure}
\includegraphics[angle=270, scale=0.7]{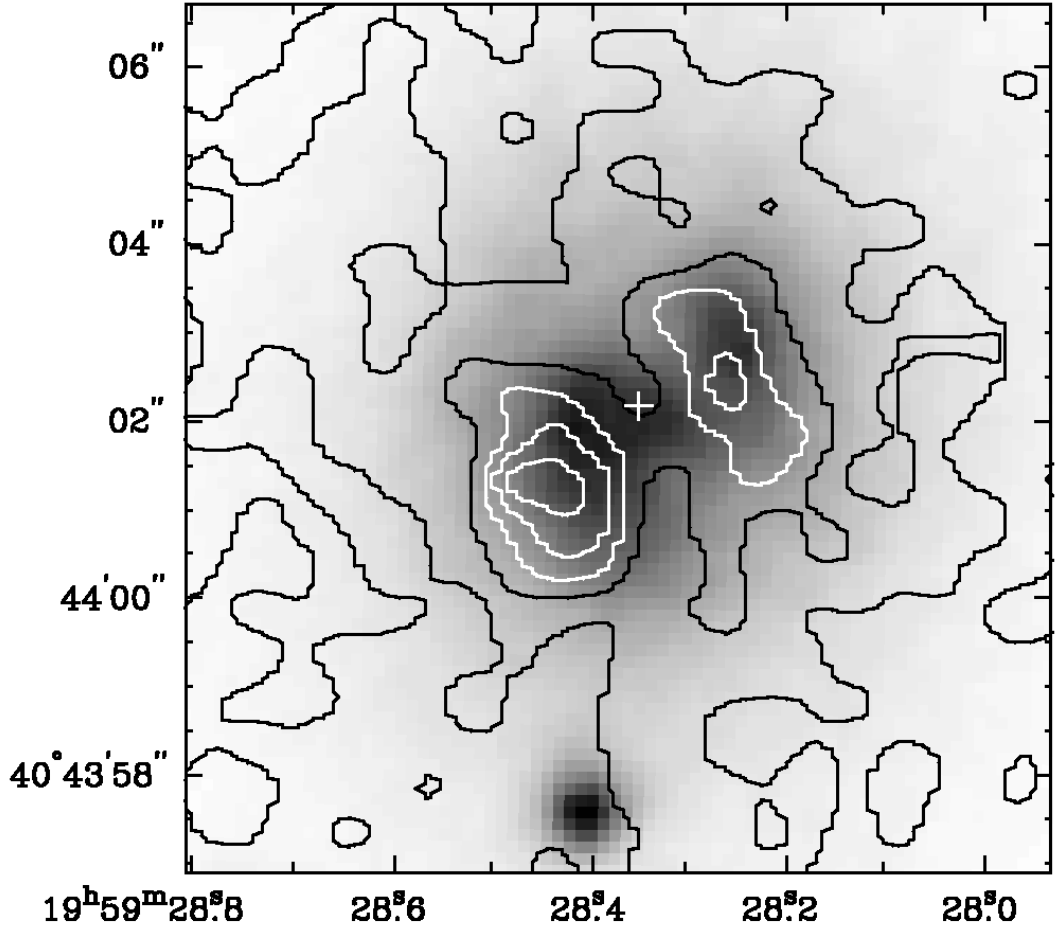}

\caption{ \label{fig:opt_r_band} A superposition of the \emph{Chandra} 3.2 s
  frame-time data with events in the range 0.25 to 1.00 keV (contours) on an
  optical r band image taken through a filter with center wavelength/bandwidth
  7531/361 \AA\ (grey scale; Stockton et al. 1994). The FWHM seeing for the
  optical image was 0\farcs81. The contours of the \emph{Chandra} image are at
  2, 6, 10, 14, and 18 cts ($0\farcs5$ pixel)$^{-1}$. The \emph{Chandra} image
  has been resampled to ten times smaller pixel size and then smoothed by a
  Gaussian of FWHM $0\farcs5$. The grey scale of the optical image is
  proportional to the intensity squared between $5 \times 10^{-20}$ (white) and
  $6 \times 10^{-19} \ergpcmsqpspA$ (black).}

\end{figure}

We recomputed the plate solutions of the infrared K$^\prime$ and optical r band
images of Stockton et al. (1994) using the known positions of stars in the
field of Cygnus A (Griffin 1963). Unsaturated stars c, d, g and h (using the
nomenclature of Griffin 1963) were used for the K$^\prime$ band and c, d and g
for the r band. Having obtained the plate solution using the stars, the {\sc
iraf} tool {\sc center} places the K$^\prime$ nucleus at $\alpha_{{\rm
K}^\prime} = 19^{\rm h}\ 59^{\rm m}\ 28\fs378$, $\delta_{{\rm K}^\prime} =
40^\circ\ 44^\prime\ 02\farcs09$ (J2000), which is within $0\farcs4$ of the
radio core and represents coincidence given the errors. The nucleus in the r
band image is not point-like, so cannot be simply aligned with the radio core.
Instead, the same offset (less than $0\farcs4$) was applied to the r band image
as was applied to the K$^\prime$ image. The r band image shows bright peaks of
emission at the nucleus, and to the northwest and southeast. The bright,
extended optical emission correlates well with the extended X-ray emission (see
Fig.~\ref{fig:opt_r_band}).

\begin{figure}
  \includegraphics[angle=0, scale=0.80]{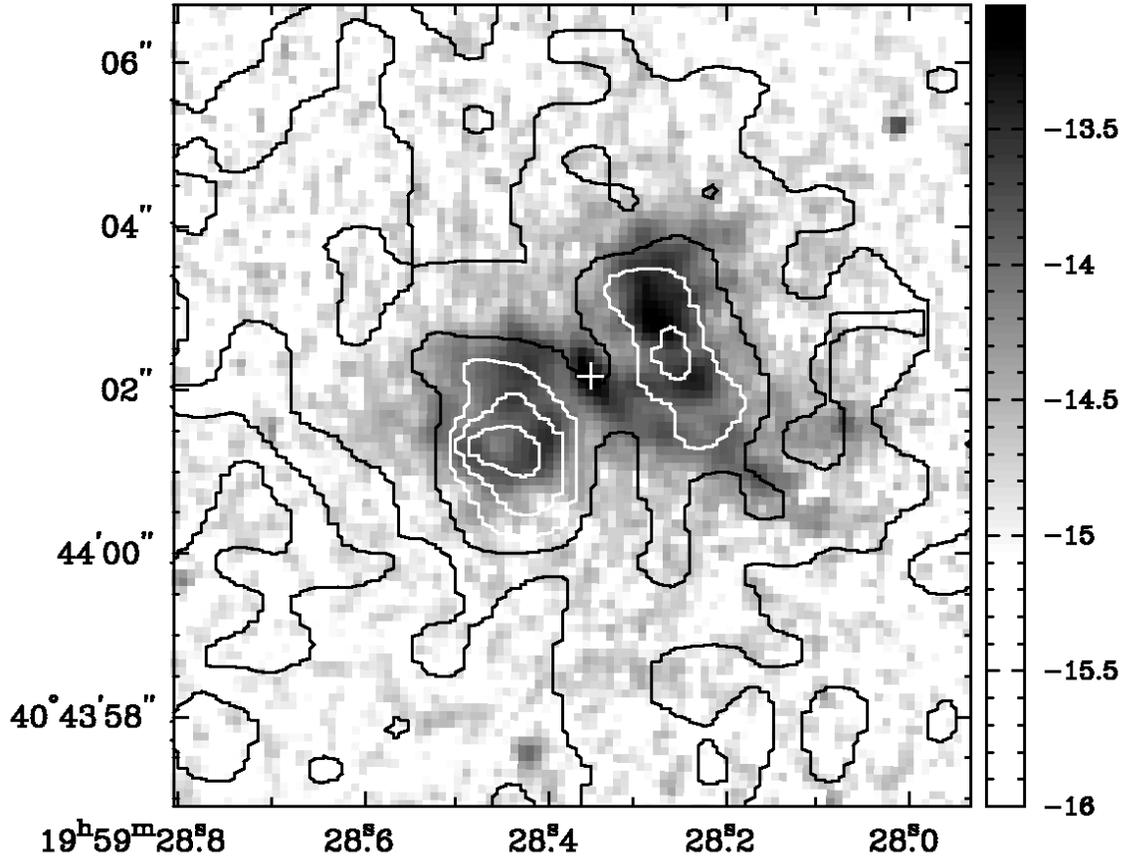}

\caption{  \label{fig:hst_oiii} A superposition of the \emph{Chandra} 3.2 s
  frame-time data with events in the range 0.25 to 1.00 keV (solid contours) on
  an HST image taken through a Linear Ramp Filter at redshifted [\ion{O}{3}]
  $\lambda$5007 (grey scale). The peaks of the 2.00 to 7.50 keV X-ray emission
  and the central optical component of the HST images were aligned with the
  radio core (see Section \ref{sec:hst_align}). The contours are at 2, 6, 10,
  14, and 18 cts ($0\farcs5$ pixel)$^{-1}$. The \emph{Chandra} image has been
  resampled to ten times smaller pixel size and smoothed by a Gaussian of FWHM
  $0\farcs5$. The optical image has approximately 0\farcs1 pixel$^{-1}$. The
  grey scale of the optical image is proportional to the logarithm of the line
  intensity and ranges from $1 \times 10^{-16}$ (white) to $9 \times 10^{-14}
  \ergpcmsqps ({\rm arcsec})^{-2}$ (black), as indicated by the vertical bar.}

\end{figure}

\begin{figure}
  \includegraphics[angle=270, scale=0.7]{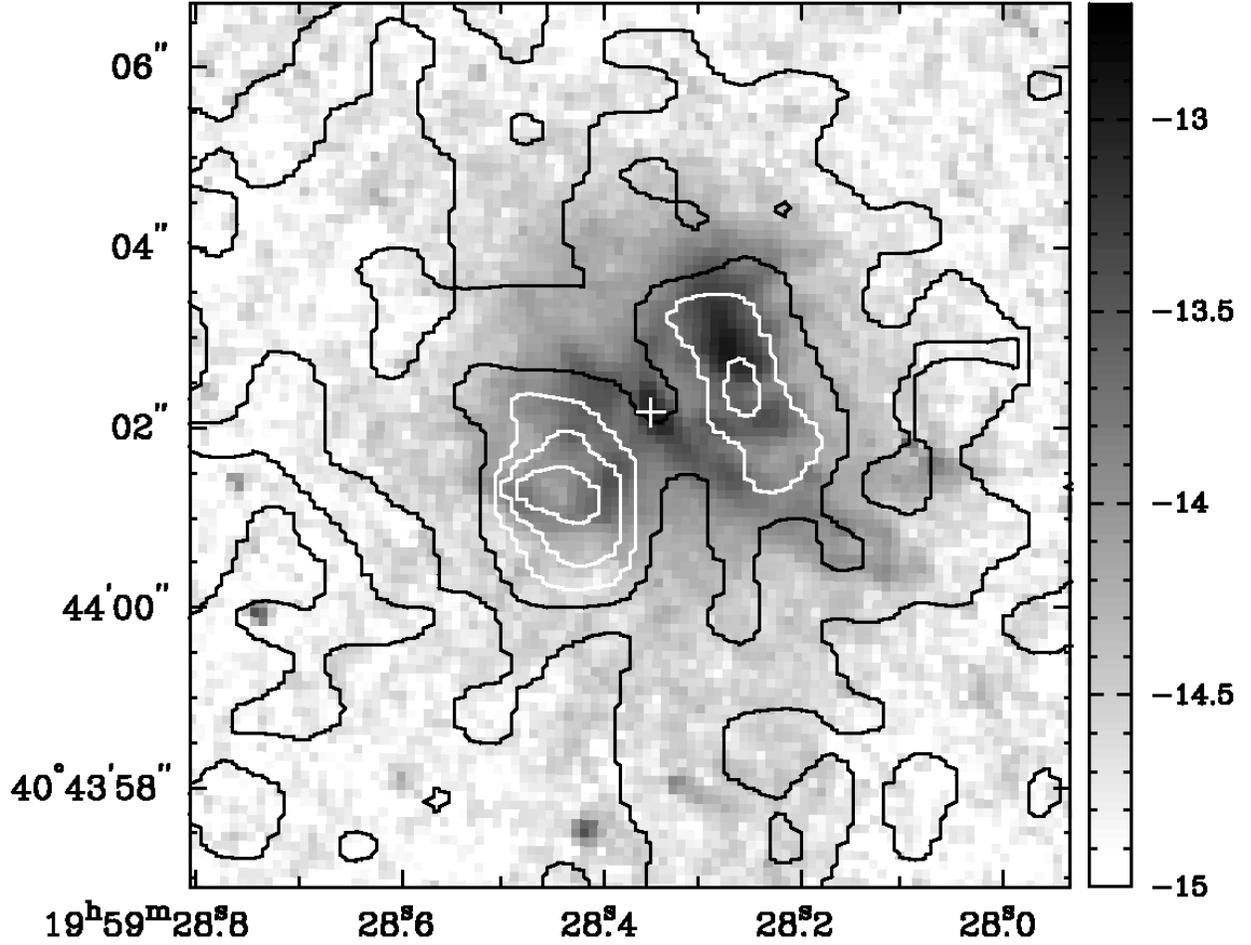}

\caption{ \label{fig:hst_halpha} A superposition of the \emph{Chandra} 3.2 s
  frame-time data with events in the range 0.25 to 1.00 keV (solid contours) on
  an HST image taken through a Linear Ramp Filter at redshifted H$\alpha$ and
  [\ion{N}{2}] $\lambda \lambda 6548$, 6583 line emission (grey scale). The
  peaks of the 2.00 to 7.50 keV X-ray emission and the central optical
  component of the HST images were aligned with the radio core
  (Section~\ref{sec:hst_align}). The contours are at 2, 6, 10, 14, and 18 cts
  ($0\farcs5$ pixel)$^{-1}$. The \emph{Chandra} image has been resampled to ten
  times smaller pixel size and then smoothed by a Gaussian of FWHM $0\farcs5$.
  The optical image has approximately 0\farcs1 pixel$^{-1}$. The grey scale of
  the optical image is proportional to the logarithm of the intensity and
  ranges from $1 \times 10^{-15}$ (white) to $2 \times 10^{-13} \ergpcmsqps
  ({\rm arcsec})^{-2}$ (black), as indicated by the vertical bar.}

\end{figure}

HST imaging in wavebands containing optical emission lines (Jackson et al.
1994; Jackson et al 1996) reveals three distinct structures in a roughly
$4^{\prime\prime} \times 4^{\prime\prime}$ region around the nucleus. There are
two ``clouds'', one to the northwest and one to the southeast, and a ``bar'' connecting the S
extreme of the northwest cloud to the N extreme of the southeast cloud. In the middle of the
``bar'' there is a bright optical knot at $\alpha_{\rm HST} = 19^{\rm h}\
59^{\rm m}\ 28\fs344$, $\delta_{\rm HST} = 40\degmark\ 44^\prime\ 2\farcs25$
(J2000) (Jackson et al. 1994), which is within 0\farcs1 of the radio core.
Thus, in aligning the optical, radio and X-ray images, the central optical
component in the HST images was assumed to coincide with the radio and X-ray
nuclei. There are bright peaks in [\ion{O}{3}] $\lambda 5007$ and H$\alpha$ +
[\ion{N}{2}]$\lambda\lambda 6548$, 6583 emission coincident with the hard X-ray
nucleus, and the northern region of the extended X-ray emission to the northwest (see
Figs. \ref{fig:hst_oiii} and \ref{fig:hst_halpha}). Diffuse [\ion{O}{3}] line
emission is seen from an approximately bi-conical region centered on the
nucleus and in this regard the X-ray and [\ion{O}{3}] emissions appear well
correlated (as also found in NGC 1068 [Young, Wilson \& Shopbell 2001], where
the ionization of the X-ray emitting gas is dominated by photoionization
[Paerels et al. 2000]). \label{sec:hst_align}

\section{Discussion}

\subsection{The Hidden Nucleus} \label{sec:hidden_nuc}

The \emph{Chandra} observations of the nucleus of Cygnus A identify the source
of hard X-ray emission with an unresolved point source located within
$0\farcs4$ of, and probably coincident with, the radio core. The spectrum of
this point source is well described by a power law of photon index $\Gamma =
1.5^{+0.12}_{-0.12}$, with rest-frame 2 -- 10 keV luminosity (corrected for
absorption) of $L_X ({\rm nuc}) = 3.7 \times 10^{44} \ergps$ obscured by a
column density of $2.0^{+0.1}_{-0.2} \times 10^{23} \pcmsq$. The power law
slope has been determined with unprecedented precision for Cygnus A, and is at
the lower end of the distribution of photon indices for NLRG (see Fig. 2(a) of
Sambruna et al. 1999). A narrow Fe K$\alpha$ line is observed in the energy
range of \ion{Fe}{2} -- \ion{Fe}{18}, with an equivalent width of approximately
200~eV. The fact that the hard X-ray power law is detected out to 100 keV
confirms the similarity to Seyferts and quasars. However, we find no evidence
for broad Fe K$\alpha$.

The above noted value of $L_X ({\rm nuc})$ in the 2 -- 10 keV band and the
[\ion{O}{3}]$\lambda5007$ luminosity ($2.6\times10^{43} \ergps$, after
correction for 1.5 mag of Galactic extinction -- see Osterbrock \& Miller 1975)
are in accordance with the X-ray -- [\ion{O}{3}] luminosity correlation for
Seyfert galaxies and unobscured radio galaxies (e.g., Mulchaey et al. 1994,
Sambruna et al. 1999), and are comparable to the two type 2 QSO candidates IRAS
23060+0505 and IRAS 20460+1925 (see Fig. 5 in Halpern et al. 1998). The X-ray
to far infrared luminosity ratio in Cygnus A $L_{\rm X} (2 - 10 \keV) / L_{\rm
FIR} = 0.18$ and is similar to QSOs (e.g., Elvis et al. 1994). The far infrared
luminosity was calculated from the \emph{IRAS} fluxes using $F_{\rm FIR} = 1.26
\times 10^{-14} (2.58S_{60} + S_{100})$ W m$^{-2}$, where $S_{60}$ and
$S_{100}$ are the flux densities at 60 $\mu$m and 100 $\mu$m, respectively, in
units of Jy.

We attempt to estimate the nuclear optical absolute magnitude which cannot be
measured directly because of the heavy extinction toward the nucleus. First, we
assume the spectral energy distribution (SED) of Cygnus A is the same as the
median SED of radio-loud quasars presented in Elvis et al. (1994). Then,
normalizing to the X-ray flux, the absolute magnitude in the B band is found to
be $M_{\rm B}=-22.4$. Alternatively, the H$\beta$ luminosity -- $M_{\rm B}$
correlation for type 1 Seyfert galaxies and QSOs (Ho \& Peng 2001), gives
$M_{\rm B}=-21.2$. Note, however, that the H$\beta$ luminosities adopted in Ho
\& Peng (2001) are a combination of the broad and narrow components and the
broad component dominates in type 1 QSOs.  We used only the narrow component of
H$\beta$ for Cygnus A and therefore this last value of $M_{\rm B}$ should be
regarded as an upper limit (i.e. a lower limit on the B band luminosity). This
upper limit is consistent with the estimate obtained using the SED.

\subsection{Electron-scattered Nuclear Flux} \label{sec:electron-scattered}

In this section we discuss the possibility that the extended bi-polar soft
X-ray continuum emission to the northwest and southeast of the nucleus is
electron-scattered nuclear light. In such a model the continuum spectrum of the
electron-scattered component is identical to that of the nucleus, and the
observations are consistent with the directly viewed hard component and the
soft component having the same photon index (i.e. $\Gamma_h = \Gamma_l$,
Sections \ref{sec:nuc_spec} and \ref{sec:nw_se_spec}). An extrapolation of the
soft X-ray power law has an unabsorbed rest-frame 2 -- 10 keV luminosity of
approximately $3.7 \times 10^{42} \ergps$ (see Section \ref{sec:nw_se_spec})
which corresponds to approximately 1\% of the unobscured 2 -- 10 keV nuclear
luminosity ($3.7 \times 10^{44} \ergps$; see Section \ref{sec:nuc_spec}). This
is compatible with the range of scattering fractions inferred in Seyfert 2
galaxies, in which typically 0.25 -- 5\% of the nuclear light is hypothesized
to be scattered into our line of sight (Turner et al. 1997b), and less than the
5 -- 10\% inferred in those objects exhibiting broad polarized H$\alpha$ (Awaki
et al. 2000); of course, there may be, in addition, intrinsically extended
thermal emission. If the X-rays are scattered by an optically thin population
of electrons in a bi-conical region, then \begin{eqnarray*} L_{\rm scattered} &
= & L_{\rm intrinsic} \frac{\Omega}{4\pi} \tau_{\rm scattering} \\ & = & L_{\rm
intrinsic} \left( 1 - \cos \theta \right) \tau_{\rm scattering} \end{eqnarray*}
where $\theta$ is the opening half-angle of the bi-conical region, observed by
HST to be $\theta \simeq 55\degmark$. Using the luminosities given above, we
find an optical depth through one cone of $\tau_{\rm scattering} \simeq 0.02$
and a column density $N_e ({\rm scattering}) = \tau_{\rm scattering} / \sigma_T
= 3.5 \times 10^{22} \pcmsq$. This is significantly higher than both the
Galactic column density to Cygnus A, $N_H ({\rm Gal}) = 3.3 \times 10^{21}
\cmsq$, and the upper limit to the intrinsic column density to the southeast or
northwest regions, $\max \{ N_H({\rm SE}), N_H({\rm NW}) \} - N_H({\rm Gal})
\approxlt 2.9 \times 10^{21} \pcmsq$ (see Section \ref{sec:column}). The lack
of strong absorption in the \emph{Chandra} soft X-ray spectrum rules out such a
high column of \emph{cold}, ionized gas.

As just noted, the observed intrinsic absorbing column density to the soft
X-ray emission, assuming the absorbing gas is cold, is $N_H ({\rm observed,\
intrinsic}) \approxlt 2.9 \times 10^{21} \pcmsq$ (see above), while the column
density for the scattering medium is inferred to be much larger, $N_H ({\rm
scattering}) \simeq 3.5 \times 10^{22} \pcmsq$. This discrepancy may be
reconciled if the scattering medium is ionized and hence has a lower opacity to
soft X-rays. Ideally, the absorber should be modeled as an ionized gas.
However, we can get an idea of the conditions required by writing
\begin{eqnarray*} N_H ({\rm observed,\ intrinsic}) = N_H ({\rm scattering})
\frac{ \sigma_{\rm abs} ({\rm ionized}) }{ \sigma_{\rm abs} ({\rm cold}) },
\end{eqnarray*} where $\sigma_{\rm abs} ({\rm ionized})$ and $\sigma_{\rm abs}
({\rm cold})$ are the absorption cross sections of ionized and cold gas,
respectively, and bearing in mind that $\sigma_{\rm abs} ({\rm ionized}) /
\sigma_{\rm abs} ({\rm cold})$ is a function of energy even for a single
ionization parameter for the ionized gas. The absorbing column intrinsic to
Cygnus A affects the observed spectrum most in the range $\simeq 0.5$ -- 1 keV.
At just above $E = 0.5 \keV$, $\sigma_{\rm abs} ({\rm cold}) \simeq (1 - 2)
\times 10^{-22} / (E/\keV)^3 \cmsq$ (Morrison \& McCammon 1983), and
$\sigma_{\rm abs} ({\rm ionized}) \simeq (2 - 3) \times 10^{-23} / (E/\keV)^3
\cmsq$ (Krolik \& Kallman 1984, their Fig. 4) for a photoionized gas with an
ionization parameter (taking into account their different definition of the
ionization parameter) of $\xi = L_X ({\rm nuc}) / (n_H R^2) \simeq 1$ erg cm
s$^{-1}$, where $n_H$ is the number density of hydrogen nuclei and $R$ is the
distance from the nucleus. This ionization parameter corresponds to a plasma
temperature of $2 \times 10^4 \K$ and such a plasma has almost an order of
magnitude lower soft X-ray opacity than neutral gas near 0.5 keV. Similarly, at
$E = 1 \keV$ $\sigma_{\rm abs} ({\rm cold}) \simeq 2.4 \times 10^{-22} \cmsq$
and $\sigma_{\rm abs} ({\rm ionized}) \simeq 3.6 \times 10^{-23} \cmsq$ for a
photoionized gas with $\xi = 63$ erg cm s$^{-1}$ and $T = 9.1 \times 10^4 \K$
(again from Krolik \& Kallman 1984, their Fig. 4). Thus, such a plasma has
almost an order of magnitude lower soft X-ray opacity than neutral gas, but
this time at 1 keV. Thus a conservative lower limit is $\xi > 1$ erg cm
s$^{-1}$; it is a lower limit because there could be \emph{cold} gas present
with the intrinsic column ($N_H ({\rm observed,\ intrinsic}) \approxlt 2.9
\times 10^{21} \pcmsq$) and \emph{zero} absorption from the scattering column,
which would then have to have a very high value of $\xi$.

The measured unabsorbed X-ray luminosity of the nucleus then implies $n_e R^2
\approxlt 3.7 \times 10^{44} (\xi / 1 $ erg cm s$^{-1})^{-1}$ (taking $n_e
\simeq n_H$). The column density of the scattering electrons $N_e ({\rm
scattering}) \simeq 3.5 \times 10^{22} \pcmsq$ corresponds to a number density
of $n_e \simeq 11 / R (\kpc) = 6 \pcmcu$ for the observed $R = 1.9$ kpc.
Combining the equations for $n_e R^2$ and $n_e R$ we find $R < 3.4 (\xi / 1$
erg cm s$^{-1})^{-1} \kpc$, which agrees with the observed distance of the soft
X-ray emission from the nucleus if $\xi \simeq 1$ erg cm s$^{-1}$. A low value
of $\xi$ is consistent with the low ionization state of the Fe K$\alpha$ line
observed from the nucleus and circumnuclear region (see
Tables~\ref{tab:nuc_lines} and \ref{tab:nw_se_lines}).

The presence of highly ionized Ne lines in the soft X-ray spectrum may be used
to constrain the location of gas emitting those lines. To photoionize Ne to
\ion{Ne}{7} and \ion{Ne}{10} requires $20 \approxlt \xi \approxlt 300$ erg cm
s$^{-1}$ (Kallman \& McCray 1982) implying $n_e R^2 \approxlt 1.9 \times
10^{43} (\xi / 20$ erg cm s$^{-1})^{-1}$. If $n_e = 6 \pcmcu$, we get $R \le
569 \pc$, substantially smaller than the observed nuclear distance of the soft
X-ray peaks (Figs.~\ref{fig:chandra_6cm}~--~\ref{fig:hst_halpha}), which are
dominated by continuum radiation. It is possible, however, that $n_e$ may be
lower, allowing a larger value of $R$. This would be the case in a multi-phase
gas, e.g., with warm dense blobs embedded in a hotter less dense gas, which
would be responsible for the observed highly ionized line emission.

The nuclear absolute magnitude is estimated by Ogle et al. (1997) using the
optical polarized flux as \begin{eqnarray*} M = M_{\rm PF} - 2.5\log \left(
\frac{1}{q\tau P} \right) - A, \end{eqnarray*} where $M_{\rm PF}$ is the
absolute magnitude of the polarized flux, $q$ is the source covering fraction,
$\tau$ is the scattering optical depth, $P$ is the intrinsic polarization of
the scattered light, and $A$ is the extinction. The observed $M_{\rm PF} =
-15.2$ (rest B band corrected for Galactic extinction, western component only,
$H_0$ = 50 km s$^{-1}$ Mpc$^{-1}$), $P=11$\%, and $q=0.86$\% (Ogle et al. 1997)
gives $M_{\rm B} = -22.8 - 2.5\log \tau - A_{\rm B}$. Our estimation of $M_{\rm
B}$ from the SED and this relation suggest that $\tau$ is order of 1 if the
extinction is small. This $\tau$ is much larger than that estimated from the
extended soft X-ray emission if the scattering particles are electrons,
suggesting instead that the polarized optical light is scattered by dust,
unless the value of $\Omega$ for the X-rays is much smaller than we assumed.

\subsection{A Non-Scattering Origin for the Extended Soft X-ray Flux}

The extended, soft X-ray emission may also be described by a thermal
bremsstrahlung model (see Table~\ref{tab:nw_se_spec}). The direct
bremsstrahlung emission from a plasma with the density inferred in the
scattering model ($n_e \sim 6 \pcmcu$; Section \ref{sec:electron-scattered})
would dominate over the scattered flux unless the emitting and scattering
medium has a temperature below $10^6$ K, which is significantly lower than that
required in the bremsstrahlung model -- $kT = 1.5^{+1.5}_{-0.5} \keV \simeq 2
\times 10^7 \K$. An argument against the extended emission being entirely due
to a thermal plasma is the extremely low metallicity required to provide a good
description of the data. A bremsstrahlung model has no metal lines and the best
fitting {\sc mekal} model has $Z = 0.03 Z_\odot$. This is unrealistic in what
is apparently a metal rich environment (see Paper III).

\subsection{Conclusions}

We have obtained the highest resolution ($< 1\arcsec$) image to date of the
X-ray emission of the nuclear region of Cygnus A with \emph{Chandra} and also
made quasi-simultaneous \emph{RXTE} observations. To summarize, we find:

\begin{itemize}

\item The hard X-ray nucleus is located within 0\farcs4 of the radio core (i.e.
coincident to within the current \emph{Chandra} systematics), and unresolved.
The nucleus is well described by a power law of photon index $\Gamma_h \simeq
1.5$ heavily absorbed by a column density of $N_H \simeq 2 \times 10^{23}
\pcmsq$. The rest-frame 2 -- 10 keV unabsorbed luminosity of the nucleus is
$L_X ({\rm nuc}) \simeq 3.7 \times 10^{44} \ergps$. Emission lines from H-like
and/or He-like Ne, highly ionized Si and ``neutral'' Fe are observed from the
nucleus. A strong Fe K absorption edge is seen, confirming the heavy
obscuration of the nucleus.

\item The \emph{RXTE} observations show that the spectrum of the cluster gas
may be described by a bremsstrahlung model with temperature $kT \simeq 6.9$~keV
and a rest-frame 2 -- 10 keV unabsorbed luminosity of $L_X ({\rm cluster}) =
1.2 \times 10^{45} \ergps$. Emission lines from ionized Fe K$\alpha$, and
either, or both of, Ni K$\alpha$ and Fe K$\beta$ are also observed from the
cluster.

\item Bi-polar soft X-ray emission is observed to extend $\simeq 1\farcs2$
towards the northwest and southeast of the nucleus of Cygnus A, and aligns well with the
bi-polar structures seen in optical continuum and line radiation. The bi-polar
morphology may be produced or enhanced by the passage of a dust lane across the
nucleus, providing an additional absorbing column density of $N_H ({\rm cold\
gas\ in\ dust\ lane}) \simeq 2.4 \times 10^{21} \pcmsq$ and suppressing the
very soft X-ray flux. Emission lines from \ion{Ne}{7} -- \ion{Ne}{9},
\ion{Si}{2} -- \ion{Si}{11}, and tentatively H-like Ne are observed in the soft
X-ray spectrum of the northwest and southeast regions.

\item Extrapolation of the best power law model of the soft (0.5 -- 2 keV),
extended, circumnuclear emission to higher energies gives a 2 -- 10 keV
luminosity of $\simeq 3.7 \times 10^{42} \ergps$, which is 1\% of that of the
unabsorbed nuclear luminosity in the same band. Furthermore, the photon index
($\Gamma_l$) of this soft emission agrees with that ($\Gamma_h$) of the
directly viewed hard X-ray emission, and so the soft emission is consistent
with being electron-scattered X-rays from the nucleus. The scattering region
must be ionized to ionization parameter $\xi \approxgt 1$ in order to be
sufficiently transparent to soft X-rays. The column density of the scattering
electrons is inferred to be much lower than that required to generate the
polarized optical light by electron scattering, suggesting the optical light
is, in fact, scattered by dust.

\item The extended soft X-ray emission is well correlated with ionized gas
observed optically with HST, in particular images in [\ion{O}{3}] $\lambda
5007$ and H$\alpha$ + [\ion{N}{2}] $\lambda\lambda 6548$, 6583. This result is
consistent with the soft X-ray emitting gas being photoionized.

\end{itemize}

Future high spectral resolution observations with the \emph{Chandra} and
\emph{XMM-Newton} gratings would be able to resolve triplet lines, such as
\ion{Ne}{9}, that are good diagnostics of the plasma temperature and the
relative importance of photoionization and collisional ionization. In addition,
the presence of narrow radiative recombination continua would confirm that the
soft X-ray emitting gas is photoionized (cf. Paerels et al. 2000).

We are grateful to Patrick Shopbell for assistance with the HST images, and
thank A. Stockton for providing electronic versions of his r and K$^\prime$
band images. The VLA radio map was provided in electronic form by R. A. Perley.
We thank the referee, Dan Harris, for constructive criticism that improved the
clarity and presentation of this paper. YT is supported by the Japan Society
for the Promotion of Science Postdoctoral Fellowship for Young Scientists. This
research was supported by NASA through grants NAG 81027 and NAG 81755.


\singlespace


\onecolumn


\vfil\eject\clearpage
\begin{deluxetable}{ccccccc}
\tablewidth{0pt} \rotate
  
\tablecaption{Previous X-ray Spectral Models of the Nucleus and Cluster Gas of
Cygnus A \label{tab:previous_xray}}

\tablecolumns{7}

\tablehead{ \colhead{Observatory} & \colhead{Reference} & \colhead{$N_H$ (nuc)}
& \colhead{$\Gamma$ (nuc)} & \colhead{$L_X ({\rm nuc})$ \tablenotemark{c}} &
\colhead{$kT$ (cluster)} & \colhead{$L_X ({\rm cluster})$ \tablenotemark{c}} \\
\colhead{} & \colhead{} & \colhead{[\pcmsq]} & \colhead{} & \colhead{[\ergps]}
& \colhead{[\keV]} & \colhead{[\ergps]}}

\startdata

\emph{EXOSAT} / \emph{HEAO1} & Arnaud et al. (1987) & $8.2^{+10.0}_{-3.0}
\times 10^{22}$ & 1.7\tablenotemark{a}~~ (1.78\tablenotemark{b}~ ) & $5.3
\times 10^{44}$ & $4.1^{+5.6}_{-1.5}$ & $6.7 \times 10^{44}$ \\

\emph{Ginga} & Ueno et al. (1994) & $3.75^{+0.75}_{-0.71} \times 10^{23}$ &
$1.98^{+0.18}_{-0.20}$ & $1.1 \times 10^{45}$ & $7.3^{+1.8}_{-1.3}$ & $1.2
\times 10^{45}$ \\

\emph{ROSAT} HRI & Harris et al. (1994) & $8.2 \times 10^{22}$\tablenotemark{a}
& & $3.6 \times 10^{45}$ & & \\

\emph{ASCA} & Sambruna et al. (1999) & $1.1^{+2.1}_{-0.6} \times 10^{23}$ &
$1.80^{+0.28}_{-0.43}$ & $2.5 \times 10^{44}$ & $7.6^{+1.5}_{-1.5}$ & $9.9
\times 10^{44}$ \\

& & $1.5^{+2.1}_{-0.9} \times 10^{23}$ & $1.76^{+0.35}_{-0.34}$ & $7.9 \times
10^{44}$ & $8.4^{+1.3}_{-1.4}$ & $8.9 \times 10^{44}$ \\

\enddata

\tablenotetext{a}{Parameter fixed.}

\tablenotetext{b}{Value if allowed to vary.}

\tablenotetext{c}{Over energy range 2 -- 10 keV.}

\end{deluxetable}


\vfil\eject\clearpage
\begin{deluxetable}{cccc}
\tablewidth{0pt} 
  
\tablecaption{Spectral Models of the Nucleus and Cluster Gas of Cygnus A
\label{tab:nuc_spec}}

\tablecolumns{4}

\tablehead{ \colhead{Model component} & \colhead{Parameter} &
  \colhead{Model 1} & \colhead{Model 2}}

\startdata

Column density to low energy emission (nucleus) & $N_{\rm H}$ [$\pcmsq$] &
$1.9^{+1.7}_{-1.0} \times 10^{21}$ & $3.3 \times 10^{21}$ \tablenotemark{a} \\

\\

Low energy ($< 2 \keV$) bremsstrahlung (nucleus) & $kT$ [keV] & &
$6.6^{+\infty}_{-5.2}$\\

& $K_{\rm Brem}\tablenotemark{e}$ & & $9.5^{+4.8}_{-1.3} \times 10^{-5}$ \\

\\

Low energy ($<2 \keV$) power-law (nucleus) & $\Gamma_l$ &
$1.52^{+0.12}_{-0.12}$ \tablenotemark{f} & \\

& $K_{\rm PL}$\tablenotemark{c} & $6.0^{+1.7}_{-1.1} \times 10^{-5}$ & \\

\\

Column density to nucleus & $N_{\rm H}$ [$\pcmsq$] & $ 2.0^{+0.1}_{-0.2} \times
10^{23}$ & $2.0^{+0.1}_{-0.2} \times 10^{23}$ \\

\\

High energy power-law (nucleus) & $\Gamma_h$ & $1.52^{+0.12}_{-0.12}$ &
$1.53^{+0.12}_{-0.15}$ \\

& $K_{\rm PL}$\tablenotemark{c} & $5.2^{+1.6}_{-1.1} \times 10^{-3}$
& $5.3^{+1.6}_{-1.0} \times 10^{-3}$ \\

\\

Bremsstrahlung (cluster gas)\tablenotemark{b} & $kT$ [keV] &
$6.9^{+0.3}_{-1.0}$ & $6.8^{+0.5}_{-0.8}$ \\

& $K_{\rm Brem}$\tablenotemark{e} & $2.3^{+0.5}_{-0.3} \times
10^{-2}$ & $2.3^{+0.4}_{-0.3} \times 10^{-2}$ \\

\\

Relative normalizations & \emph{Chandra} & 1.00 & 1.00 \\

& \emph{RXTE} PCA & $1.31^{+0.75}_{-0.26}$ & $1.34^{+0.39}_{-0.31}$ \\

& \emph{RXTE} HEXTE & $1.08^{+0.31}_{-0.07}$ & $1.10^{+0.51}_{-0.11}$ \\

\\

Fit statistic & \chisq\ / d.o.f. & 181 / 185 & 177 / 184 \\

\enddata

\tablenotetext{a}{Parameter fixed at Galactic column density.}

\tablenotetext{b}{The cluster gas contributes significantly to only
	the \emph{RXTE} data sets.}

\tablenotetext{c}{$K_{\rm PL} = \phpcmsqps\keV^{-1}$ at 1
  keV.}

\tablenotetext{d}{$K_{\rm Fe} = \phpcmsqps$ in the line.}

\tablenotetext{e}{$K_{\rm Brem} = 3.02 \times 10^{-15} \int n_e n_I dV / (4 \pi
D_A^2)$, where $n_e$ is the electron density ($\pcmcu$), $n_I$ is the ion
density ($\pcmcu$), and $D_A$ is the angular size distance to the source (cm).}

\tablenotetext{f}{Fixed to the same value as found for $\Gamma_h$.}

\end{deluxetable}


\vfil\eject\clearpage \begin{deluxetable}{ccccccccc}
\tabletypesize{\footnotesize} \tablewidth{0pt} \rotate

\tablecaption{Emission Lines From the Nucleus (\emph{Chandra}, Aperture
Diameter 2\farcs5) and Cluster Gas (\emph{RXTE}, Spatial Resolution $\simeq
1\degmark$)\label{tab:nuc_lines}}

\tablecolumns{9}

\tablehead{ \colhead{Model} & \colhead{Location} & \colhead{} & 
        \colhead{Energy} & \colhead{Line} & \colhead{} & \colhead{Rest-frame
        energy} & \colhead{$K$\tablenotemark{a}} & \colhead{EW} \\ \colhead{} &
        \colhead{} & \colhead{} & \colhead{[keV]} & \colhead{} & \colhead{} &
        \colhead{[keV]} & \colhead{} & \colhead{[eV]}}

\startdata

1 & Nucleus & & 0.90 -- 1.02 & \ion{Ne}{9} triplet -- \ion{Ne}{10} & &
$0.97^{+0.03}_{-0.06}$ & $6.0^{+9.6}_{-4.9} \times 10^{-6}$ &
90\tablenotemark{b} \\

1 & Nucleus & & 1.76 -- 1.86 & \ion{Si}{7} -- \ion{Si}{13} & &
$1.83^{+0.02}_{-0.06}$ & $3.7^{+2.3}_{-2.3} \times 10^{-6}$ &
148\tablenotemark{b} \\

1 & Nucleus & & 6.40 -- 6.44 & \ion{Fe}{2} -- \ion{Fe}{18} & &
$6.40^{+0.05}_{-0.03}$ & $5.9^{+1.4}_{-1.8} \times 10^{-5}$ &
182\tablenotemark{c} \\

1 & Cluster gas & & 6.65 -- 6.69 & \ion{Fe}{24} -- \ion{Fe}{25} & &
$6.70^{+0.04}_{-0.04}$ & $2.9^{+0.9}_{-0.6} \times 10^{-4}$ &
444\tablenotemark{d} \\

1 & Cluster gas & $\left\{ \begin{array}{c} \\ \\ \end{array} \right.$ &
\begin{tabular}{c}7.70 -- 8.08 \\ 7.68 -- 7.88 \end{tabular} &
\begin{tabular}{c} \ion{Ni}{25} -- \ion{Ni}{28} \\ \ion{Fe}{22} K$\beta$ --
\ion{Fe}{25} K$\beta$ \end{tabular} & $\left. \begin{array}{c} \\ \\
\end{array} \right\}$ & $7.90^{+0.21}_{-0.22}$ & $5.0^{+0.9}_{-1.5} \times
10^{-5}$ & 115\tablenotemark{d} \\

\\

\cline{1-9}

\\

2 & Nucleus & & 0.90 -- 0.92 & \ion{Ne}{9} triplet & &
$0.92^{+0.07}_{-0.02}$ & $1.7^{+0.6}_{-1.2} \times 10^{-5}$ &
202\tablenotemark{e} \\

2 & Nucleus & & 1.76 -- 1.86 & \ion{Si}{7} -- \ion{Si}{13} & &
$1.83^{+0.03}_{-0.07}$ & $3.3^{+2.7}_{-2.4} \times 10^{-6}$ &
109\tablenotemark{e} \\

2 & Nucleus & & 6.40 -- 6.44 & \ion{Fe}{2} -- \ion{Fe}{18} & &
$6.41^{+0.04}_{-0.04}$ & $6.6^{+0.9}_{-2.4} \times 10^{-5}$ &
202\tablenotemark{c} \\

2 & Cluster gas & & 6.65 -- 6.69 & \ion{Fe}{24} -- \ion{Fe}{25} & &
$6.70^{+0.05}_{-0.05}$ & $2.7^{+0.4}_{-0.7} \times 10^{-4}$ &
434\tablenotemark{d} \\

2 & Cluster gas & $\left\{ \begin{array}{c} \\ \\ \end{array} \right.$ &
\begin{tabular}{c} 7.70 -- 8.08 \\ 7.74 -- 7.88 \end{tabular} &
\begin{tabular}{c} \ion{Ni}{25} -- \ion{Ni}{28} \\ \ion{Fe}{23} K$\beta$ --
\ion{Fe}{25} K$\beta$ \end{tabular} & $\left. \begin{array}{c} \\ \\
\end{array} \right\}$ & $7.90^{+0.19}_{-0.17}$ & $4.8^{+1.3}_{-1.2} \times
10^{-5}$ & 115\tablenotemark{d} \\

\enddata

\tablenotetext{a}{$K = \phpcmsqps$ in the line.}

\tablenotetext{b}{With respect to the unabsorbed low-energy power-law
component.}

\tablenotetext{c}{With respect to the unabsorbed high-energy
power-law component.}

\tablenotetext{d}{With respect to the unabsorbed bremsstrahlung
component modeling the cluster emission.}

\tablenotetext{e}{With respect to the unabsorbed low-energy
bremsstrahlung component.}

\end{deluxetable}


\vfil\eject\clearpage
\begin{deluxetable}{cccccccc}
\tablewidth{0pt} \rotate
  
\tablecaption{Spectral Models of the Northwest and Southeast Circumnuclear Regions (Two
  Sectors, Diameter 7\arcsec) \label{tab:nw_se_spec}}

\tablecolumns{8}

\tablehead{ \colhead{Frame-time} & \colhead{Model} & \colhead{$N_{\rm H}$} &
\colhead{$kT$} & \colhead{Abundance} & \colhead{$\Gamma_l$} &
\colhead{Normalization\tablenotemark{a}} & \colhead{$\chisq$ / d.o.f.} \\
\colhead{[sec]} & \colhead{} & \colhead{[\pcmsq]} & \colhead{[keV]} &
\colhead{[$\times Z_\odot$]} & \colhead{} & \colhead{} & \colhead{}}

\startdata

0.4 & {\sc mekal} & $3.3 \times 10^{21}$ \tablenotemark{b} &
$4.9^{+5.9}_{-1.9}$ & 1.00\tablenotemark{c} & & $K_{\rm MEKAL} =
3.7^{+0.6}_{-0.6} \times 10^{-4}$ & 50 / 27 \\

0.4 & {\sc mekal} & $3.3 \times 10^{21}$ \tablenotemark{b} &
$1.5^{+1.5}_{-0.5}$ & $0.03^{+0.24}_{-0.03}$ & & $K_{\rm MEKAL} =
6.4^{+2.6}_{-2.0} \times 10^{-4}$ & 44 / 26 \\

0.4 & Brem & $3.3 \times 10^{21}$ \tablenotemark{b} &
$1.5^{+1.5}_{-0.5}$ & & & $K_{\rm Brem} = 2.1^{+0.9}_{-0.5} \times
10^{-4}$ & 44 / 27 \\

0.4 & PL & $3.3 \times 10^{21}$ \tablenotemark{b} & & & $2.17^{+0.36}_{-0.38}$
& $K_{\rm PL} = 1.4^{+0.1}_{-0.2} \times 10^{-4}$ & 42 / 27 \\

0.4 & PL + lines & $3.3 \times 10^{21}$ \tablenotemark{b} & & &
$2.01^{+0.46}_{-0.54}$ & $K_{\rm PL} = 1.1^{+0.2}_{-0.2} \times 10^{-4}$ & 22 /
21 \\

\enddata

\tablenotetext{a}{Model normalizations are\\ $K_{\rm MEKAL} = 10^{-14} / (4 \pi
[D_A \{ 1+z \} ]^2 ) \int n_e n_H dV$, where $n_e$ is the electron density
($\pcmcu$), $n_H$ is the hydrogen density ($\pcmcu$) and $D_A$ is the angular
size distance to the source (cm),\\ $K_{\rm Brem} = 3.02 \times 10^{-15} \int
n_e n_I dV / (4 \pi D_A^2)$, where $n_e$ is the electron density ($\pcmcu$),
$n_I$ is the ion density ($\pcmcu$), and $D_A$ is the angular size distance to
the source (cm),\\ $K_{\rm PL} = \phpcmsqps\keV^{-1}$ at 1 keV.}

\tablenotetext{b}{Fixed at the Galactic column.}

\tablenotetext{c}{Parameter fixed.}

\end{deluxetable}


\vfil\eject\clearpage
\begin{deluxetable}{cccccc}
\tablewidth{0pt} \rotate
  
\tablecaption{Emission Lines From the Northwest and Southeast Circumnuclear Regions (Two
Sectors, Diameter 7\arcsec) \label{tab:nw_se_lines}}

\tablecolumns{6}

\tablehead{ \colhead{Frame-time} & \colhead{Energy} & \colhead{Line} &
  \colhead{Rest-frame energy} & \colhead{$K$\tablenotemark{a}} & \colhead{EW}
  \\ \colhead{[sec]} & \colhead{[keV]} & \colhead{} & \colhead{[keV]} &
  \colhead{} & \colhead{[eV]}}

\startdata

0.4 & 0.89 -- 0.92 & \ion{Ne}{7} -- \ion{Ne}{9} & $0.91^{+0.01}_{-0.02}$ &
$1.8^{+1.0}_{-0.8} \times 10^{-5}$ & 127 \\

0.4 & 1.02 & \ion{Ne}{10} Ly $\alpha$ ? & $1.08^{+0.28}_{-0.09}$ &
$5.5^{+5.1}_{-5.5} \times 10^{-6}$ & 54 \\

0.4 & 1.74 -- 1.82 & \ion{Si}{2} -- \ion{Si}{11} &
$1.76^{+0.06}_{-0.03}$ & $4.2^{+3.0}_{-2.7} \times 10^{-6}$ & 110 \\

0.4 & 6.40 -- 6.44 & \ion{Fe}{2} -- \ion{Fe}{18} & $6.39^{+0.03}_{-0.03}$ &
$5.0^{+0.9}_{-1.8} \times 10^{-5}$ & 226 \\

\enddata

\tablenotetext{a}{$K = \phpcmsqps$ in the line.}

\end{deluxetable}


\vfil\eject\clearpage

\end{document}